\documentclass{aa}

\usepackage{graphicx}
\usepackage{txfonts}
\usepackage{siunitx}
\usepackage{xcolor}

\begin{document}

   \title{The effect of Jupiter on the CAI storage problem}

   \author{S. Jongejan\inst{1}
          \and
          C. Dominik\inst{1}
          \and
          C.P. Dullemond \inst{2}
          }

   \institute{Anton Pannekoek Institute for Astronomy, University of Amsterdam, Science Park 904, 1098XH Amsterdam, The Netherlands
   \and
   Institute for Theoretical Astrophysics, Center for Astronomy, Heidelberg University, Albert-Ueberle-Str. 2, 69120 Heidelberg, Germany
   }

  \abstract
   {Meteorites preserve an imprint of conditions in the early Solar System. By studying the distribution of calcium-aluminium-rich inclusions (CAIs) that are embedded within meteorites, we can learn about the dynamical history of the protoplanetary disk from which our Solar System formed. A long-standing problem concerning CAIs is the CAI storage problem. CAIs are thought to have formed at high temperatures near the Sun, but they are primarily found in carbonaceous chondrites, which formed much further out, beyond the orbit of Jupiter. Additionally, radial drift of CAI particles should have removed them from the solar protoplanetary disk several million years before the parent bodies of meteorites in which they are encountered would have accreted.}
   {We revisit a previously suggested solution to the CAI storage problem by Desch, Kalyaan, and Alexander which proposed that CAIs were mixed radially outward through the disk and subsequently got trapped in a pressure maximum created by Jupiter's growing core opening a planet gap. Our aim is to investigate whether their solution still works when we take into account the infall phase during which the disk builds up from the collapse of a molecular cloud core.}
   {We build a 1D numerical code in Python using the DISKLAB package to simulate the evolution of the solar protoplanetary disk, starting with a collapsing molecular cloud. CAIs are created in the model by thermal processing of solar nebula composition dust, and subsequently transported through the disk by turbulent diffusion, radial drift and advection by the gas.}
   {We find that outward transport of CAIs during the infall phase is very efficient, possibly mixing them all the way into the far outer disk. Subsequent inward radial drift collects CAIs in the pressure maximum beyond Jupiter's orbit while draining the inner disk, roughly reproducing parts of the result by Desch et al. By introducing CAI formation so early, abundances out to 100 AU remain significant, possibly not consistent with some meteoritic data. It is possible to create a disk that does not expand as far out and also does not push CAIs as far out by using a very slowly rotating cloud.}
   {}


   \maketitle
   
\section{Introduction}

A wealth of information about the conditions in the early Solar System can be obtained through the study of meteorites. These celestial objects are thought to have undergone little change since their formation over 4.5 billion years ago, and as such they preserve an imprint of the conditions in the solar protoplanetary disk. The distribution of calcium-aluminium-rich inclusions (CAIs) in present-day meteorites is one such piece of information that can tell us something about the dynamical history of our Solar System. These structures were not expected to survive long enough to end up in any meteorite parent bodies at all, yet somehow they must have survived the harsh conditions at the time of their birth for long enough to do so. To add to this mystery, they are also predominantly found in locations far away from the Sun, in whose vicinity they must have originally formed. These two issues together constitute the CAI storage problem. In this paper we present the results of a numerical model of the early Solar System, which attempts to explain the observed distribution of CAIs in meteorites. Our model builds on the solution previously proposed by Desch et al. (2018). That paper attempts to construct a detailed, fine-tuned model for the formation and outward mixing of CAIs, with the goal to fit and address many aspects of the record that is available to us in the form of meteoritic data. While meteoritic data samples limited regions of the Solar System, it is very detailed and produces an important guide on how to model processes in the young Solar System. Our goal in the present study is not to attempt a reconstruction of the Solar System and the meteoritic record on a similar level of detail as Desch et al. Instead, we perform a calculation with similar assumptions about physical processes in the disk, in particular with similar assumptions about the disk viscosity and the CAI-forming processes, but start the computation earlier by computing the effects of the infall phase from a molecular cloud core. We show that the infall phase can significantly influence the "starting conditions" in the final phase of the circumstellar disk in which the Desch model is anchored.

In this introduction we start by giving a brief overview about the different kinds of meteorites encountered. We then describe CAIs and the CAI storage problem in more detail, before discussing some of the work previously done on this topic.

Meteorites are an important tool for understanding the history of the Solar System and planet formation mechanisms, since they preserve information about the conditions in the solar protoplanetary disk, such as the abundances of their constituent elements, isotope ratios, or temperatures they must have been subjected to. The classification of meteorites into different classes, clans, groups, and subgroups can be a murky topic upon which there is no universal agreement. Here we only present a simplified overview of the different kinds of meteorites. An extensive modern overview of meteorite classification can for example be found in Weisberg et al (2006).

Broadly speaking, we can distinguish two types of meteorites: achondrites, which are composed of material that underwent melting or differentiation inside a massive parent body, and chondrites, which did not undergo such processes. Using this definition, achondrites encompass not only rocky achondrites, which represent crustal material, but also iron meteorites, which are composed of an iron-nickel alloy thought to come from the core of a differentiated parent object. An intermediate case exists in the form of pallasites, which consist of silicates embedded within an iron matrix, appearing to have formed from a mix of core as well as mantle or crustal material. Another category of achondrite are the primitive achondrites, which only underwent partial melting (making them more closely related to chondrites), or which perhaps did melt, but without solidifying in crystalline form as a result of it. Other than achondrites originating from parent asteroids or comets, a small number of finds are thought to actually have originated on Mars or the Moon.

Chondrites make up the majority of the known meteorite population. Because chondrites did not undergo melting after accreting into a parent body, they are the most primitive kind of meteorite, most closely preserving an imprint of the conditions in the solar protoplanetary disk. There are three main classes of chondrites: ordinary chondrites (OCs), which are the most common type, enstatite chondrites (ECs), which are rich in the mineral enstatite, and carbonaceous chondrites (CCs), which have relatively high carbon abundances. Some smaller classes also exist, as well as ungrouped individual meteorites that might represent the first kind of an entirely new class. Each of the main chondrite classes can be further subdivided into various groups, which are generally thought to sample separate meteorite parent bodies. A defining feature of chondrites is that chondrules are found embedded within them. Chondrules are millimeter-sized spherical droplets of igneous rocks that must have melted at high temperatures before they ended up inside their parent meteorites. This is a first hint that material that was processed at high temperatures must somehow have been transported out to colder regions of the solar protoplanetary disk (Brownlee et al. 2006).

The formation time of chondrite parent bodies can be determined indirectly in several ways. Radiometric dating of components such as chondrules, which must have formed before the parent body itself, provides upper limits on the age of the parent body. Similarly, lower limits can be determined through radiometric dating of minerals that are known to have formed only after the parent body itself accreted. Finally, estimates of the maximum temperature reached by the parent body provide model-dependent limits on the accretion time. Using these constraints, enstatite chondrite parent bodies are estimated to have formed first, around 1.8 Myr after the formation of the Solar System, followed by the ordinary chondrite parent bodies after around 2.1 Myr. Carbonaceous chondrite parent bodies formed last. The accretion time of the parent bodies for the different groups the CCs are subdivided in is more spread out over time, but ranges from 2.4 to 4.1 Myr after the formation of the Solar System (Desch et al. 2018). Constraints on meteorite ages also make it possible to put constraints on the timing of planet formation processes.

Different kinds of chondrites can be spectroscopically matched to certain groups of asteroids that are found in the asteroid belt today. This way, it can also be determined roughly what heliocentric distance these meteorites originated from. Asteroids associated with enstatite chondrites are found closest to the Sun, around 2 AU. Those linked to ordinary chondrites dominate the region between 2 and 2.5 AU, while the asteroids similar to carbonaceous chondrites are commonly found between 2.5 and 4 AU (Desch et al. 2018). While this does not necessarily mean that the meteorite parent bodies also originally accreted at these locations, at least this ordering of chondrite types by current heliocentric distance seems to match that of their original formation locations. Evidence for this is for example provided by their water content. While enstatite chondrites contain virtually no water, ordinary chondrites do contain small amounts, and also show evidence for aqueous alteration. Carbonaceous chondrites on the other hand are water-rich (Alexander et al. 2012). Since condensation of water-rich minerals is only possible in the relatively cool region of the disk beyond the snow line, this indicates that carbonaceous chondrites formed relatively far from the Sun.

Warren (2011) discovered that meteorites can be divided into two distinct clusters based on their isotopic abundances. Carbonaceous chondrites are enriched in certain neutron-rich isotopes, such as $^{50}$Ti or $^{54}$Cr, whereas ordinary and enstatite chondrites, as well as most achondrites (collectively referred to as the non-carbonaceous or NC meteorites), are relatively depleted in these isotopes. This NC-CC dichotomy implies that the two clusters of meteorites formed from two isotopically distinct reservoirs of material. Furthermore, because the parent bodies of the meteorites in which this dichotomy is observed are known to have accreted several million years apart, these two reservoirs must also have coexisted for several million years without significant mixing of material between them. A likely explanation for the separation of the two reservoirs is the formation of Jupiter, which opened a planet gap in the solar protoplanetary disk that prevented material exchange between the reservoirs (Kruijer et al. 2017). A good candidate to explain where the isotopic differences between the NC and CC reservoirs originated from in the first place is heterogeneities in the Solar System's parent molecular cloud. It is possible this cloud was not homogeneous in composition, or that its composition changed over time. Because infall from the molecular cloud affects different regions of the protoplanetary disk in different ways, it is possible that material containing neutron-rich isotopes was primarily added to the CC reservoir, or that material depleted in these isotopes was added primarily to the NC reservoir (Nanne et al. 2019). Indeed, the presence of daughter isotopes of $^{26}$Al in meteorites, a radioactive isotope with a half-life of only 0.72 Myr, is evidence that the Solar System's molecular cloud must have been recently polluted with radioactive isotopes, possibly originating from a supernova or Wolf-Rayet stellar winds, something that could explain heterogeneities in either space or time.

The present-day coexistence of asteroids linked to both NC and CC meteorites in the asteroid belt can be explained as a result of Jupiter's migration after the asteroid parent bodies accreted, first in- and then outward. This would scatter the different asteroid populations, after which they ended up together in the same asteroid belt (Walsh et al. 2011).

Calcium-aluminium-rich inclusions, or CAIs, are small solid structures that are found as inclusions in certain types of meteorites. Often irregularly shaped, they range in size from micrometers to about a centimeter. They are composed of minerals such as corundum (aluminium oxide) or perovskite (calcium titanium oxide), which are enriched in refractory elements such as calcium and aluminium, and which condense at high temperatures (T $\approx$ 1400 K), implying that CAIs formed near the Sun where such temperatures would be achieved (Grossman 1972). CAIs are also the oldest dated solids, with a mean age of approximately 4567.30 $\pm$ 0.16 Myr (Connelly et al. 2012). Because of this, and because their constituent minerals are predicted to be the first to condense in a cooling gas of solar nebula composition, the age of CAIs is often equated with the age of the Solar System itself. The time of (initial) CAI formation is therefore also the time relative to which the accretion times of meteorite parent bodies were defined previously. 

CAIs are primarily found embedded in carbonaceous chondrites, in which they typically make up a couple percent of the total volume. In ordinary and enstatite chondrites on the other hand, CAI abundances are easily an order of magnitude lower, typically less than a tenth of a percent of the total volume. This distribution presents us with two serious problems that together make up the CAI storage problem. 

The first problem is that CAIs are thought to have formed at high temperatures, which are only obtained close to the Sun. As we have seen however, the carbonaceous chondrites in which they ended up formed further out in the solar protoplanetary disk, likely beyond the orbit of Jupiter in the CC reservoir. Clearly CAIs must have been transported outward through the disk to end up there, but this raises the question why did they not also end up in ordinary and enstatite chondrites, which formed at intermediate heliocentric distances. 

The second problem is related to the accretion time of carbonaceous chondrite parent bodies. Because the gas in a protoplanetary disk normally experiences an outward pressure-gradient force which partially supports it against gravitational collapse, its orbital velocity can be slightly less than Keplerian. Solid particles do not experience such a pressure-gradient force, and hence have to orbit the central star at the Keplerian velocity. The velocity difference between gas and dust leads to a drag force on the dust particles\footnote{Technically, smaller dust grains are so well coupled to the gas that they simply get dragged along with the same orbital velocity. But because this velocity is sub-Keplerian, they will drift inward all the same.} that robs them of angular momentum and causes them to slowly drift inward, spiralling towards the star over time. This is what is called radial drift of dust particles (Weidenschilling 1977). For particles the size of large CAIs (in which most of the CAI mass is contained) that form in the inner disk near the Sun, this radial drift velocity is high enough that they are expected to all vanish into the Sun on a time scale of order 10$^4$ years. This is evidently much shorter than the time CAIs would need to remain in the disk in order to be incorporated into carbonaceous chondrite parent bodies, the last of which did not accrete until over 4 Myr after CAI formation. 

The CAI storage problem can therefore be summarized as the question how CAIs managed to survive long enough in the disk to end up in any meteorites in the first place, and why they then ended up preferentially in carbonaceous chondrites, which of all meteorite types formed the furthest away from the location where CAIs themselves formed.

This problem is closely related to the problem of the NC-CC dichotomy. CAIs are enriched in many of the elements that are also found to be enriched in the CC reservoir in general. However, the NC-CC dichotomy has been found to also extend to elements which aren't present in CAIs, such as nickel (Nanne et al. 2019). This means that simply adding CAIs to a reservoir of otherwise NC composition does not lead to the CC reservoir. 

Over time, a number of mechanisms have been proposed that could explain how CAIs and other dust species can be transported outward in a protoplanetary disk. Cuzzi et al (2003) showed that inward radial drift of CAIs can be overcome by turbulent diffusion, a process in which turbulent motions of the gas within a protoplanetary disk redistribute solid particles by advection, at least part of which would be in the outward direction. This allows CAIs to survive in the disk on time scales of order 10$^6$ years. Keller and Gail (2003) performed both calculations and numerical simulations that showed that while the accretion stream in a protoplanetary disk is normally pointed inward, flows near the disk midplane can actually point outward, thereby providing another mechanism of radial outward transport of material called meridional flow. Boss et al (2012) showed that the gravitational instability, an instability that occurs when a disk becomes so massive that its own self-gravity can no longer be neglected, can lead to a rapid redistribution of mass both in- and outward. 

Cuzzi et al (2003) also proposed the CAI Factory model, which is a simplified picture of the CAI formation environment. The CAI Factory consists of a region at a nearly constant temperature, where the minerals CAIs are composed of can exist as solids, but which is too hot to allow for other silicates, such as chondrules, in solid form. The radial extent of the CAI Factory changes over time, the outer boundary moving inward as the disk cools. 

Desch, Kalyaan, and Alexander (2018) proposed a solution to the CAI storage problem, to which we will hereafter simply refer as the DKA18 model. They constructed a 1D hydrodynamics code that builds on Cuzzi's CAI Factory model as well as the suggestion of Kruijer et al (2017) that a planet gap opened by Jupiter prevented material exchange between the regions interior and exterior to the gap, previously mentioned in the context of the NC-CC dichotomy. Their simulation begins at $t = 0$ with a disk that is seeded with dust of solar nebula composition. Viscous heating creates a region near the Sun, the CAI Factory, in which the temperature reaches 1400 K. Solar nebula composition dust that enters this region is thermally processed, and a certain fraction of it is instantaneously converted into CAIs of one particular size. This conversion of dust continues for several 10$^5$ years. In the meantime, the disk is viscously evolving. CAIs are transported out of the CAI Factory, both into the Sun and outward into the disk by the effects of turbulent diffusion and meridional flow. A small part of the CAIs diffused past a heliocentric distance of 3 AU in this way, where Jupiter's core of 30 $M_{\oplus}$ is then assumed to form 0.6 Myr into the simulation. As the planet grows to its full size over the course of the next 4 Myr by accreting gas from its surroundings, it opens up a gap in the disk, where the surface density of the gas is significantly reduced. As a result of this, there exists a region just beyond the planet location where the gas density necessarily increases again in the outward direction. This means that the pressure-gradient force here points inward instead of outward as it normally does, and that gas in this region will be orbiting the Sun with super-Keplerian velocities in order to balance itself against gravitational collapse. This also reverses the sign of dust radial drift, causing CAIs to be transported outward. Some distance beyond the planet, the gas surface density and pressure reach a maximum before continuing their normal outward decrease. In this pressure bump, the gas orbits with exactly the Keplerian velocity, removing any drag force on solid particles and therefore the conditions for dust radial drift. This thus creates a situation in which CAIs in the vicinity always drift in the direction of the pressure bump, at which location they can remain in a stable orbit for millions of years, until the moment they are incorporated into the accreting meteorite parent bodies. In the meantime, CAIs in the inner disk continue to drift into the Sun unimpeded, depleting the formation location of ordinary and enstatite chondrites. At the end of the simulation, the DKA18 model predicts that all remaining CAIs in the disk are concentrated around the pressure bump behind Jupiter's orbit, where their abundance peaks at about 6\% of all solids.\footnote{Other than just calculating the abundances of CAIs and refractory elements, Desch, Kalyaan, \& Alexander also demonstrate that disk conditions consistent with the formation of chondrites matching those abundances as well as properties such as water content emerge from contextual evidence such as $^{54}$Cr anomalies and radiometric dating. They also calculate the particle size concentrated by turbulence in each chondrite's formation location and find an excellent match with observed chondrule sizes.}

While it seems that the DKA18 model conclusively solves the CAI storage problem, there are some issues with it that could have a significant impact on the results. The most important of these issues is that the model starts with a fully formed star plus disk system, within which CAI formation is then initiated. The build-up of the solar protoplanetary disk from a parent molecular cloud core is neglected. It is therefore unclear what effect the infall phase would have on the timing of CAI formation, their dynamical evolution and final abundance profile, or indeed whether the solution of Jupiter keeping CAIs in place would still work in the first place. While Yang and Ciesla (2012) did perform disk simulations in which the effects of the infall phase were taken into account, showing that CAIs that formed during the infall phase could be preserved in primitive bodies that accreted in the outer disk, they did not address the question why CAIs would be depleted in the inner disk.

\section{Methods}\label{Methods}

\subsection{DISKLAB}\label{DISKLAB}

Our model was programmed in Python using DISKLAB, a package developed by Cornelis Dullemond and Til Birnstiel\footnote{https://github.com/dullemond/DISKLAB -- User ID: dullemond \& birnstiel - Access granted upon request}, which contains many basic functions for setting up disk models, calculating basic quantities such as surface densities and temperatures, and evolving these models over time. While DISKLAB contains methods for the vertical structure of disks and for making full two-dimensional models, only one-dimensional radial (and hence rotationally symmetric) models were used for this project.

Setting up any model in DISKLAB begins by calling the DiskRadialModel class, which sets up a grid of discrete radial coordinates at which disk quantities will be calculated, as well as basic stellar parameters such as mass and luminosity, all of which can be freely modified. A surface density profile for the disk gas can then be set by hand, but usually the easier approach is to begin with a basic model such as a powerlaw disk and then modify it according to your wishes.

In addition to the (main) gas component of the disk, any number of solid components (dust species) can be added to the model. This is done by specifying a certain grain size (or alternatively, a Stokes number) and grain density, as well as the surface density for that component, which is usually the gas surface density multiplied by some dust-to-gas ratio. Dust grains of different sizes would have to be added to the model as separate components, even if they represent dust of the same chemical composition. It is not possible to follow individual dust grains through the disk with this method. 

An important property of a protoplanetary disk is its (midplane) temperature, since this directly influences the isothermal sound speed,

$$
c_s = \sqrt{\frac{k_b T}{\mu m_p}}, \eqno{(1)}
$$

\noindent (with $T$ the temperature, $k_B$ the Boltzmann constant, $\mu$ the mean molecular weight and $m_p$ the proton mass) and hence also important parameters such as the viscosity $\nu$ through

$$
\nu = \alpha c_s h .  \eqno{(2)}
$$

\noindent Here the Shakura-Sunyaev $\alpha$-parameter quantifies the strength of the angular momentum transport due to turbulence (Shakura \& Sunyaev 1973), and $h$ is the vertical scale height of the disk. The midplane temperature due to stellar irradiation $T_\text{irr}$ is calculated separately from the temperature due to viscous heating $T_\text{visc}$. The two temperatures are then combined as

$$
T = \left(T_\text{irr}^4+T_\text{visc}^4\right)^{1/4}. \eqno{(3)}
$$

\noindent The irradiation temperature itself is calculated by equating the heating rate due to irradiation by the central star $Q_\text{irr}(r)$ with the cooling rate $Q_\text{cool}(r)$:

$$
Q_\text{irr}(r) = 2 \phi(r)\frac{L_*}{4 \pi r^2}, \eqno{(4)}
$$

$$
Q_\text{cool}(r) = 2 \sigma_\text{SB} T_\text{eff}(r)^4, \eqno{(5)}
$$

\noindent with $\phi$(r) the flaring angle of the disk, $L_*$ the stellar luminosity, $\sigma_\text{SB}$ the Stefan-Boltzmann constant and $T_\text{eff}$ the effective temperature at the surface of the disk. This surface temperature is related to the midplane temperature $T_\text{irr}$ as

$$
T_\text{irr}=\left(\frac{T_\text{eff}^4}{2}\right)^{1/4}, \eqno{(6)}
$$

\noindent where the factor 2 comes from the fact that of all the stellar radiation intercepted at the disk surface, only half is re-emitted into the disk where it can heat up the midplane. (Chiang \& Goldreich 1999) (Dullemond et al. 2001) Combining Equations (4), (5) and (6) then leads to a final expression for the midplane temperature due to irradiation:

$$
T_\text{irr} = \left(\frac{\phi(r)}{2 \sigma_\text{SB}} \frac{L_*}{4 \pi r^2}\right)^{1/4}. \eqno{(7)}
$$

\noindent Similarly, the viscous temperature $T_\text{visc}$ can be found by equating the viscous heating rate $Q_\text{visc}$ with the cooling rate $Q_\text{cool}$:

$$
Q_\text{visc}(r) = \frac{9}{4} \Sigma_g(r) \nu(r) \Omega_K(r)^2, \eqno{(8)}
$$

$$
Q_\text{cool}(r) = 2 \sigma_\text{SB} T_\text{eff}(r)^4 \left(1-e^{-2 \tau_\text{Ross}}\right), \eqno{(9)}
$$

\noindent where $\Sigma_g$ is the gas surface density, $\nu$ the viscosity, $\Omega_K = \sqrt{G M_*/r^3}$ the Kepler frequency and $\tau_\text{Ross}$ the Rosseland optical depth, which depends on the dust surface density $\Sigma_d$ and Rosseland opacity $\kappa_{d,\text{Ross}}$ as

$$
\tau_\text{Ross} = \Sigma_d \kappa_{d,\text{Ross}}. \eqno{(10)}
$$

\noindent The relation used between the effective (surface) temperature $T_\text{eff}$ and the midplane temperature $T_\text{visc}$ is now

$$
T_\text{visc} = \left(\frac{1}{2} \tau_\text{Ross}+1\right)^{1/4} T_\text{eff}. \eqno{(11)}
$$

\noindent Combining Equations (8), (9), (10) and (11) then leads to an expression for the viscous temperature $T_\text{visc}$:

$$
T_\text{visc} = \left(\frac{9}{8 \sigma_\text{SB}}\frac{\frac{1}{2} \tau_\text{Ross}+1}{1-e^{-2 \tau_\text{Ross}}} \Sigma_g(r) \nu(r) \Omega_K(r)^2\right)^{1/4}. \eqno{(12)}
$$

\noindent Because this expression depends on the viscosity $\nu$ as well as the Rosseland opacity $\kappa_{d,\text{Ross}}$, which in turn depend on the temperature themselves, iteration is required. DISKLAB uses Brent's method to find the roots of this equation and solve for $T_\text{visc}$. Before this can be done however, the Rosseland mean opacities for the dust species, $\kappa_{d,\text{Ross}}$ in Equation (10), have to be specified. Ideally this is done by calculating them from the frequency-dependent opacities of the dust species that can be specified, but it is also possible to read them from a user-provided table of opacities as a function of density and temperature, to use the Bell \& Lin opacity model (Bell \& Lin 1994), or to simply specify the value to be used at each radial grid point. 

Evolving disk models over time is done by solving the time-dependent viscous disk equation: 

$$
\frac{\partial \Sigma_g}{\partial t}-\frac{3}{r}\frac{\partial}{\partial r}\left(\sqrt{r}\frac{\partial(\sqrt{r} \Sigma_g \nu)}{\partial r}\right) = \dot{\Sigma}_g, \eqno{(13)}
$$

\noindent where $\Sigma_g$ is the gas surface density, $\nu$ is the viscosity and $\dot{\Sigma}_g$ is a source term for the gas surface density that could correspond to for example infall or photoevaporation. This is a diffusion equation, which DISKLAB can solve using an implicit integration scheme. A consequence of this is that the solutions should be fairly accurate even for large time steps, while this would lead to large errors using an explicit method. It does remain true that multiple smaller time steps lead to more accurate results than a single large one. By default, the boundary condition for Equation (13) is that the gradient of $\Sigma_g$ vanishes at the inner boundary, though this can be changed to instead set it to some custom value. 

The evolution of dust components is handled separately. The dynamics of a dust particle are determined by its frictional stopping time $\tau_\text{stop}$, caused by the aerodynamic drag on the particle, and the related Stokes number, which is a dimensionless constant describing how well the dust is coupled to the gas in the disk. The expression for this stopping time first of all depends on the mean free path of gas molecules $\lambda_\text{mfp}$, which is calculated as

$$
\lambda_\text{mfp} = \frac{1}{n_\text{gas} \sigma_{H_2}}, \eqno{(14)}
$$

\noindent with $\sigma_{H_2} =  2\cdot10^{-15} \text{ cm}^2$  and the gas number density $n_\text{gas}$ 

$$
n_\text{gas} = \frac{\rho_\text{gas}}{\mu m_p}, \eqno{(15)}
$$

\noindent where $\rho_\text{gas}$ is the gas density, $\mu = 2.3$ is the mean molecular weight and $m_p$ the proton mass. There are two different physical regimes for the drag force on a solid particle, depending on the grain size $a$ compared to the mean free path $\lambda_\text{mfp}$. In the Epstein regime, which holds when $a < 9/4 \lambda_\text{mfp}$, the grain size is small compared to the mean free path, so the fluid is basically a collisionless collection of molecules following a Maxwell velocity distribution. In this regime, the stopping time takes the form

$$
\tau_\text{stop} = \frac{\xi a}{\rho_\text{gas} v_\text{th}}, \eqno{(16)}
$$

\noindent where $\xi$ is the (interior) grain density and $v_\text{th}$, the thermal velocity of the gas particles, is given by

$$
v_\text{th} = \sqrt{\frac{8 k_B T}{\pi \mu m_p}} = \sqrt{\frac{8}{\pi}} c_s. \eqno{(17)}
$$

\noindent In the Stokes regime, which holds when $a \geq 9/4 \lambda_\text{mfp}$, the dust particles are relatively large, and the gas flows around them as a fluid. In this regime, the precise form of the stopping time depends on the Reynolds number:

$$
\text{Re} = \frac{2 a \Delta v}{v_\text{mol}}, \eqno{(18)}
$$

\noindent where $\Delta v$ is the velocity difference between the gas and the dust in the disk, and the molecular viscosity $v_\text{mol}$ is given by

$$
v_\text{mol} = \frac{1}{2} v_\text{th} \lambda_\text{mfp}. \eqno{(19)}
$$

\noindent When Re $<$ 1: (Birnstiel et al. 2010)

$$
\tau_\text{stop} = \frac{2 \xi a^2}{9 v_\text{mol} \rho_\text{gas}}. \eqno{(20)}
$$

\noindent When $1 < \text{Re} < 800$: (Perets \& Murray-Clay 2011)

$$
\tau_\text{stop} = \frac{8 \xi a}{3 C_D \rho_\text{gas} \Delta v}, \eqno{(21)}
$$

\noindent where the drag coefficient $C_D$ is given by 

$$
C_D = \frac{24}{\text{Re}} \left(1+0.27 \text{Re}\right)^{0.43} + 0.47\left(1-e^{-0.04\text{Re}^{0.38}}\right). \eqno{(22)}
$$

\noindent When Re $>$ 800: (Birnstiel et al. 2010)

$$
\tau_\text{stop} = \frac{6 \xi a}{\rho_\text{gas} \Delta v}. \eqno{(23)}
$$

\noindent Regardless of which form for the stopping time applies to some particle, the Stokes number is then calculated as

$$
\text{St} = \Omega_K \tau_\text{stop}.  \eqno{(24)}
$$

\noindent With the Stokes number determined, we can then see how the dust evolves over time. First of all, as a result of the drag force on the dust caused by the velocity difference $\Delta v$ with the gas, the dust undergoes radial drift with a velocity $v_d$ given by

$$
v_d = \frac{1}{1+\text{St}^2}\left(v_r+\text{St}\frac{c_s^2}{\Omega_K r}\frac{d \text{ln} p}{d \text{ln} r}\right), \eqno{(25)}
$$

\noindent where the last term represents the pressure gradient in the gas, and the radial velocity of the gas itself $v_r$ is given by

$$
v_r = - \frac{3}{\sqrt{r} \Sigma_g} \frac{\partial (\sqrt{r} \Sigma_g \nu)}{\partial r}. \eqno{(26)}
$$

\noindent The full time-dependent evolution of the dust includes both this radial drift and mixing due to the turbulent motions in the gas, which drags the dust along:

$$
\frac{\partial \Sigma_d}{\partial t} + \frac{1}{r} \frac{\partial \left(r \Sigma_d v_d\right)}{\partial r} - \frac{1}{r} \frac{\partial}{\partial r} \left(r D_d \Sigma_g \frac{\partial}{\partial r}\left(\frac{\Sigma_d}{\Sigma_g}\right)\right) = \dot{\Sigma}_d. \eqno{(27)}
$$

\noindent Here $\Sigma_d$ is the dust surface density and $\dot{\Sigma}_d$ is a source term, analogously to Equation (13) for the gas. $D_d$ is a diffusion coefficient given by

$$
D_d = \frac{1}{\text{Sc}}\frac{1}{1+\text{St}^2} \nu, \eqno{(28)}
$$

\noindent where the Schmidt number Sc is defined as

$$
\text{Sc} = \frac{\nu}{D_g}, \eqno{(29)}
$$

\noindent where $D_g$ is the gas turbulent diffusivity. Just as Equation (13), Equation (27) is a diffusion equation that is solved in DISKLAB using implicit integration, which means it should be stable even when larger time steps are used. 

Importantly for this project, DISKLAB can also include the effects of infall from a molecular cloud into the model. To this end, it follows Hueso \&  Guillot (2005) in combining the models of Shu (1977) and Ulrich (1976), which handle the radial infall and the effect of rotation, respectively. This model starts with a molecular cloud core of mass $M_c$ that is assumed to be isothermal with temperature $T_c$, spherically symmetric with radius $R_c$, and rotating as a solid body with rotation rate $\Omega_c$. This cloud then starts to collapse from the inside out as an expansion wave propagates outward with the sound speed $c_s$, causing every shell of mass it passes through to collapse onto the star+disk system. The mass accretion rate is assumed to remain constant during this phase:

$$
\dot{M} = 0.975 \frac{c_s^3}{G}, \eqno{(30)}
$$

\noindent with G the gravitational constant. Material falling onto the disk this way always ends up within the centrifugal radius $r_\text{c}(t)$, which is given by

$$
r_c(t) = \frac{r_\text{shell}(t)^4 \omega(r_\text{shell})^2}{G M(t)}, \eqno{(31)}
$$

\noindent where $r_\text{shell}(t)$ is the distance from the center of the molecular cloud of the shell where the expansion wave passes at time $t$, $\omega(r_\text{shell})$ the angular momentum contained in that shell, and $M(t)$ the total mass that has been accreted onto the star+disk from the cloud up until that time. Since both $r_\text{shell}$ and $M(t)$ are proportional to time (due to the constant sound speed and accretion rate), $r_c \propto t^3$ and infalling matter ends up progressively further away from the star. The way this matter is spread out over the disk is then

$$
\dot{\Sigma}_g(r,t) = \frac{\dot{M}}{\pi r_c^2}\frac{1}{8}\left(\frac{r}{r_c}\right)^{-3/2}\left[1-\left(\frac{r}{r_c}\right)^{1/2}\right]^{-1/2}. \eqno{(32)}
$$

\noindent This is the $\dot{\Sigma}_g$ that enters Equation (13) for the time evolution of the gas as the source term. The source term for the dust, $\dot{\Sigma}_d$ in Equation (27), can then simply be found by multiplying $\dot{\Sigma}_g$ by the dust-to-gas ratio.

Especially in a model that includes infall, it is important to ensure that the disk does not become so massive that its own self-gravity starts to play a role, leading to the gravitational instability. This would produce non-axially symmetric effects (spiral waves) in the disk, which can't be properly treated with the axially symmetric models DISKLAB produces. The stability of the disk against this phenomenon can be checked in DISKLAB with a simple command that calculates the Toomre $Q$ parameter (Toomre 1964) as

$$
Q = \frac{c_s \Omega_K}{\pi G \Sigma_g}. \eqno{(33)}
$$

\noindent The disk remains stable as long as $Q > 2$. Under this condition, pressure forces and shear can act faster to destroy overdensities than self-gravity can produce them.

In practice, operations in DISKLAB are performed by applying functions to the DiskRadialModel object that is created at the start. This way, many individual calculations can be performed using only a single line of code. It is then up to the user to combine the different functionalities into a sensible model. Evolving a model over time can be done using a for-loop, where each iteration corresponds to the next time step, and within which the individual functions are called to solve the viscous disk equation and update other time-dependent parameters such as temperatures, velocities and masses. Because any parameter that can vary with radius in the disk is stored as a Python array, with each entry corresponding to one point on the chosen radial grid, they can also easily be manipulated by hand, using standard Python commands. It is for example possible to simulate a chemical reaction in which one type of dust is transformed into another by removing surface density from the array for the first dust species and adding it to that of the second.

Only minor changes were made to the standard DISKLAB code itself. Two new functions were added to introduce a planet and open a gap in the disk, in addition to the existing gap models in the package. These will be described in the next section. The radial velocity of the dust was also set to equal that of the gas at the innermost three grid points only, overriding Equation (25) there, due to a problem with a boundary condition that will also be described in the next section.

\subsection{Model setup}\label{Model Setup}

\begin{table*}[t]
\centering
\begin{tabular*}{\textwidth}{|c @{\extracolsep{\fill}} c c c c c|}
\hline
Population & $a$ & $\xi$ & Dust-to-gas (cloud) & Processed dust fraction & Name \\
\hline
1 & \SI{1}{\micro\metre} & 3 g cm$^{-3}$ & $1.5\cdot 10^{-3}$ & - & Solar nebula composition \\ 
2 & \SI{1}{\micro\metre} & 3 g cm$^{-3}$ & $3.5\cdot 10 ^{-3}$ & - & Solar nebula composition \\
3 & \SI{1}{\micro\metre} & 3 g cm$^{-3}$ & - & 89\% & Refractory-depleted \\
4 & \SI{600}{\micro\metre} & 3 g cm$^{-3}$ & - & 3\% & Amoeboid Olivine Aggregates (AOAs) \\
5 & \SI{2500}{\micro\metre} & 3 g cm$^{-3}$ & - & 8\% & Ca-Al-Rich Inclusions (CAIs) \\
\hline
\end{tabular*} 
\caption{Properties of the different populations of dust present in the model. Listed are the grain size $a$, the internal grain density $\xi$, the dust-to-gas ratio within the molecular cloud core, and what fraction of thermally processed population 1 dust is transformed into populations 3, 4, and 5.}
\end{table*}

The model was calculated on a 1D grid of 1000 logarithmically spaced radial points between 0.06 and 1000 AU. The inner edge of the disk $r_{\text{in}}$ at 0.06 AU is the same as used in the DKA18 model, and it seems a not unreasonable estimate for the radius of the still contracting proto-Sun.

The basic disk model used was the model by Lynden-Bell \& Pringle (1974) as described in Lodato et al (2017). But while this model is present in the code and evolving at the earliest time steps, we are really interested in letting the disk build up naturally due to the effects of infall. The initial mass was therefore chosen to have a negligibly low value, $M_0 =10^{-20} M_{\odot}$, which is about 19 orders of magnitude lower than the final disk mass. In practice then, it is not relevant what happens with this initial model at early times, because the material infalling from the molecular cloud core dominates the evolution of the disk as soon as the centrifugal radius of the infall $r_c$ exceeds the inner edge of the disk.

The way infalling matter is spread over the disk, described by Equation (32), depends solely on the properties of the molecular cloud core. The three relevant parameters are the cloud mass $M_c$, temperature $T_c$ and rotation rate $\Omega_c$.\footnote{The cloud radius $R_c$ is not an independent parameter, but can be calculated as $$R_c = \frac{G M_c}{2 c_s^2}$$ This yields a radius of roughly 0.045 pc for the chosen values for the cloud mass and temperature.} For the cloud mass, a value of $M_c = 1.05$ M$_{\odot}$ was chosen in order for the Sun to end up with roughly 1 M$_{\odot}$. The cloud temperature was set to $T_c = 14$ K, which is a typical temperature for a molecular cloud (Wilson et al. 1997). The choice for the rotation rate, $\Omega_c = 2.3\cdot10^{-14}$ rad s$^{-1}$, is more or less arbitrary, but the effect of varying this parameter will be explored later. With these choices for mass and temperature, the duration of the infall phase can be calculated by dividing the cloud mass by the accretion rate in Equation (30):

$$
t_\text{infall} = \frac{M_c}{\dot{M}} = \frac{G M_c}{0.975 c_s^3} \approx 0.4 \text{ Myr}. \eqno{(34)}
$$

Into this model, we introduced the five different species of dust from the DKA18 model. Populations 1 and 2 represent small particles ($a =$ \SI{1}{\micro\metre}) of solar nebula composition. These are present in the molecular cloud core, and as such, they are added to the disk by infall. The only difference between populations 1 and 2 is that population 1 dust can be thermally processed at temperatures of 1400 K (the CAI Factory), while population 2 dust cannot. Populations 3, 4 and 5 are the products of this thermal processing. They are not present in the molecular cloud, and as such they can only be formed from population 1 dust in the CAI Factory. The conversion of population 1 dust into the other dust species happened instantaneously at any time step when population 1 dust ended up at a radial grid point where the midplane temperature $T = 1400$ K. Population 3, which is the product of 89\% of the processed population 1 dust, consists of small grains ($a =$ \SI{1}{\micro\metre}) that are depleted in refractory elements. 3\% of the processed material turns into population 4 dust, which consists of intermediate-sized ($a =$ \SI{600}{\micro\metre}) refractory-depleted grains representing Amoeboid Olivine Aggregates (AOAs). Population 5 dust is the species we are most interested in here however. This dust, the product of 8\% of the processed population 1 dust, consists of large particles ($a =$ \SI{2500}{\micro\metre}) representing CAIs. The properties of the different dust populations are summarized in Table 1.

\begin{figure*}[t!]
  \centering
  \begin{tabular}{@{}c@{}}
    \includegraphics[width=\columnwidth]{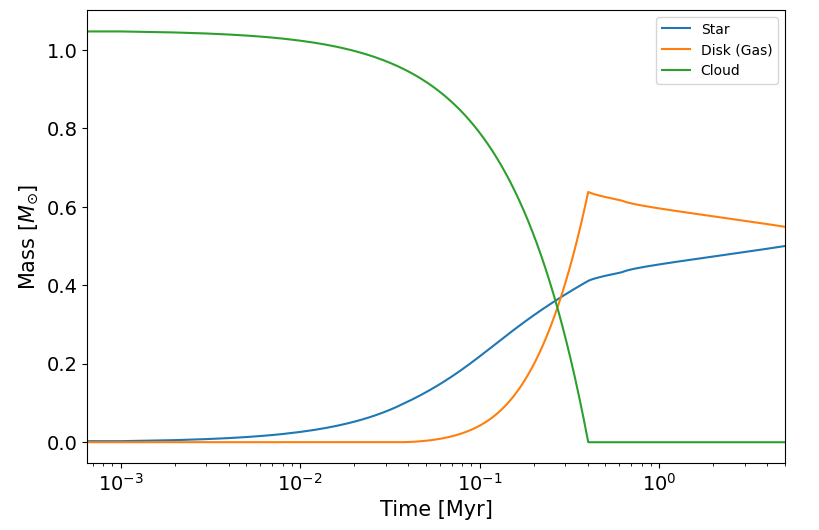}
    \includegraphics[width=\columnwidth]{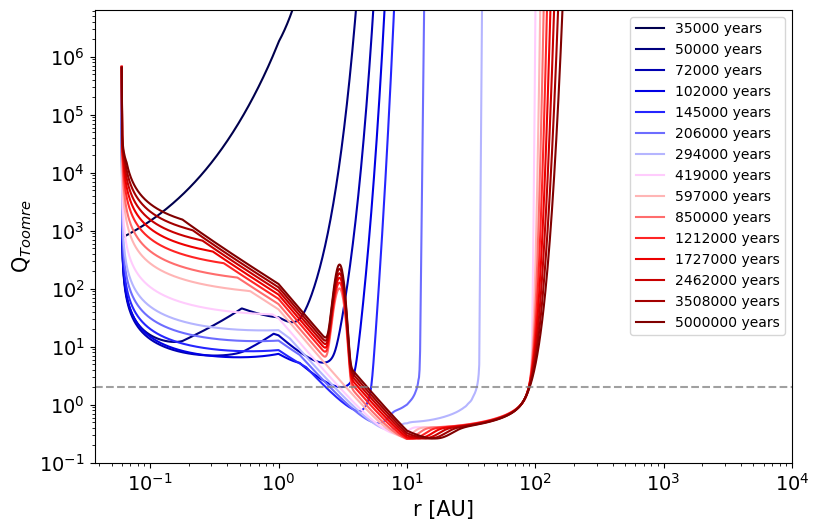}
  \end{tabular}
\caption{Result of a simulation in which the parameterization for the $\alpha$-parameter as given by Equations 36 through 38 is used throughout the infall phase. \textit{Left:} Evolution of the total mass present in each of the molecular cloud, star and disk as a function of time. In this scenario, mass can't accrete onto the star from the disk rapidly enough, so the disk keeps growing in mass until it even exceeds the mass present in the star. Even at the end of the simulation after 5 Myr, the star has only gained roughly 0.5 M$_{\odot}$ of mass. \textit{Right:} Resulting value of the Toomre $Q$ as a function of radius in the disk, shown for several intermediate time steps. As the disk grows in mass, $Q$ drops below the safe value of 2 (indicated by the horizontal dashed line) everywhere except in the innermost part of the disk. This means that the self-gravity of gas in the disk can no longer be ignored, and the disk becomes susceptible to the gravitational instability.}\label{fig:NoHighInitialAlpha}
\end{figure*}

The temperature calculation as described in Section \ref{DISKLAB} still depends on a model for the stellar luminosity $L_*$, the flaring angle $\phi(r)$ and the dust opacity $\kappa_{d,\text{Ross}}$. For the flaring angle, a global value of 0.05 radians was used. Following DKA18, the model used for the luminosity of the young Sun is that of Baraffe et al (2002):

$$
L_*(t) = 1.2 \left(\frac{t}{\text{Myr}}\right)^{-0.5} L_{\odot}. \eqno{(35)}
$$

\noindent This model is quite crude, for two main reasons. First, the $t = 0$ in the model really corresponds to the time when the Young Stellar Object (YSO) becomes visible, which should be at some point towards the end of the infall phase, when most of the envelope has dissipated. Second, the evolutionary tracks provided by Baraffe et al are quite uncertain before $t = 1$ Myr, because they depend strongly on initial conditions. In fact we can see that Equation (35) diverges towards $t = 0$. However, the crudeness of this model is not very significant in practice, because the temperature calculation, at least in the inner disk, is dominated by the viscous heating anyway. For the Rosseland opacity $\kappa_{d,\text{Ross}}$ for the dust, a constant value of 5 cm$^2$ g$^{-1}$ was chosen at every radial grid point.\footnote{It was at first attempted to calculate $\kappa_{d,\text{Ross}}$ self-consistently from the available dust components. However, while this seemed to work for a static disk model, it led to highly unstable results when evolving the model in time. For reasons not entirely clear, this caused temperatures (and hence also the sound speed $c_s$ and viscosity $\nu$) to wildly fluctuate throughout the disk, and even shoot back up towards 1400 K in the far outer disk. Clearly such a result is unphysical.} The midplane temperature anywhere in the disk was manually capped at 1400 K, the temperature at which most silicates evaporate. Further heating would lead to a decrease in dust opacity, which in turn increases the cooling rate, which causes the temperature to drop again, leading to a kind of thermostat effect. In practice, this maximum temperature was achieved through viscous heating alone during the infall phase, further decreasing the significance of the uncertain irradiation temperature. 
An important parameter we still need to specify that influences the evolution of the disk is the $\alpha$-parameter from Equation (2), which quantifies the efficiency of angular momentum transport due to turbulence in the disk. Again we follow the DKA18 model, which provides an $\alpha$-parameterization in three parts: a higher constant value in the inner disk and a lower constant value in the outer disk, with the two parts being connected by a decreasing powerlaw. In the inner disk, for $r < 1 $ AU:

$$
\alpha = 5\cdot10^{-4}. \eqno{(36)}
$$

\noindent For $1 < r < 10 \text{ AU}$:

$$
\alpha = 5\cdot10^{-4} \left(\frac{r}{\text{AU}}\right)^{-1.7}. \eqno{(37)}
$$

\noindent And for $r \geq 10$ AU:

$$
\alpha = 1\cdot10^{-5}. \eqno{(38)}
$$

\noindent The effect of different choices for the $\alpha$-value will be explored later. However, early experimentation with this model revealed that this choice of $\alpha$ is problematic during the infall phase for typical values of the cloud rotation rate $\Omega_c$. The resulting viscosity is not strong enough to rapidly remove infalling gas from the disk again, by carrying away its angular momentum and letting it accrete onto the Sun. The result of this is shown as Figure \ref{fig:NoHighInitialAlpha}. The left panel shows the mass evolution of the different system components as a function of time. Because mass cannot accrete from the disk onto the Sun rapidly enough, the disk keeps growing in mass until it eventually even exceeds the stellar mass. The right panel displays the resulting time evolution of the Toomre $Q$ parameter (Equation 33) throughout the disk, showing that this parameter drops below the safe threshold of $Q = 2$ in all but the innermost few AU of the disk. The disk therefore becomes gravitationally unstable, and the results of this model can't be trusted. Fortunately, there is a simple solution for this. We can mimic the effects of the gravitational instability by adopting an artificially enlarged $\alpha$-parameter (Armitage et al. 2001) during the infall phase. This redistributes the infalling material much more rapidly, ensuring that most of the gas flowing through the disk actually ends up in the star. It turns out a value of $\alpha = 0.6$ is sufficient to ensure that $Q > 2$ and the total disk mass $M_\text{disk} < 0.1 M_{\odot}$ at all times. This value was therefore used during the entire infall phase, after which $\alpha$ was changed to follow Equations (36-38). In addition to this, we also ran a simulation with a reduced cloud rotation rate of $\Omega_c = 1\times10^{-15}$, which is low enough that the gravitational instability never triggers, allowing us to use Equations (36-38) from the start. 

To solve Equation (13) for the viscous evolution of the gas, a boundary condition needs to be specified. The default zero-gradient boundary condition caused issues with the temperature calculation, so instead we set $\Sigma_g$ to a fixed value at the inner disk edge. A low value of $\Sigma_g = 10^{-14}$ g cm$^{-2}$ was used for the initial low-mass disk model, which was then increased to $\Sigma_g = 10$ g cm$^{-2}$ when the centrifugal radius of the infall crossed the disk inner edge, a value similar to the gas surface density predicted for the innermost disk at the end of the simulation. However, using a fixed value can cause problems with radial outflow of dust species at the inner disk edge when $\Sigma_g$ increases in the outward direction, because this creates a pressure gradient that can act as a particle trap in the same way a planet gap does. For this reason, the radial velocity of the dust species was set to equal that of the gas at the innermost three grid points only, which prevents them from getting trapped.

\begin{table*}[t]
\centering
\begin{tabular*}{\textwidth}{|p{0.2\textwidth}p{0.2\textwidth}p{0.53\textwidth}|}
\hline
Parameter & Value & Description \\
\hline
$M_c$ & 1.05 $M_{\odot}$ & Initial mass of the molecular cloud core \\
$T_c$ & 14 K & Isothermal temperature of the molecular cloud core \\
$\Omega_c$ & $2.3*10^{-14}$ s$^{-1}$ & Solid-body rotation rate of the molecular cloud core \\
$M_\text{star,0}$ & $1.05*10^{-4} M_{\odot}$ & Initial mass of the Sun \\
$M_0$ & $10^{-20} M_{\odot}$ & Mass of the initial Lynden-Bell \& Pringle disk model \\
$R_0$ & 1 AU & Truncation radius of the initial Lynden-Bell \& Pringle disk model \\
$r_\text{in}$ & 0.06 AU & Inner edge of the disk \\
$\sigma_\text{in}$ & 10 g cm$^{-2}$ & Gas surface density at the disk inner boundary \\
$\alpha_\text{in}$ & 0.6 & Global value of the $\alpha$-parameter during the infall phase \\
$t_\text{planet}$ & 0.6 Myr & Time of formation of Jupiter's core \\
$a_\text{planet}$ & 3 AU & Semi-major axis of Jupiter's orbit at the formation time \\
$m_\text{planet,0}$ & 30 $M_{\oplus}$ & Mass of Jupiter's core at the formation time \\
$\tau_\text{planet}$ & 1537.5 yr & Growth time scale of Jupiter (for post-infall time steps of 1000 years) \\
$\alpha_\text{peak}$ & 0.01 & Value of $\alpha$ at Jupiter's location after its formation \\
$\kappa_{d,\text{Ross}}$ & 5 cm$^2$ g$^{-1}$ & Global value of the dust opacity \\
\hline
\end{tabular*}
\caption{Overview of physical parameters chosen for the model. See Table 1 for the properties of the dust species. The values of $\alpha$ in the post-infall phase are given in Equations (36) to (38).}
\end{table*}

The model is now set up and ready to be evolved over time. The simulation ran for 5 million years, by the end of which CAI parent bodies are thought to have finished their formation. Since DISKLAB uses implicit integration for calculating the time evolution of the model, relatively large time steps could be employed. Unfortunately however, this does not apply to the thermal conversion of the dust species, which had to be done explicitly at each time step. We therefore used a constant time step of 0.2 years during the infall phase, when dust mixing is particularly strong, after which we switched to a time step of 1000 years for the remainder of the simulation. More information on how these time steps were chosen can be found in the Appendix. This way, each individual simulation could be finished in under a day.

\begin{figure}[t!]
  \centering
  \includegraphics[width=\columnwidth]{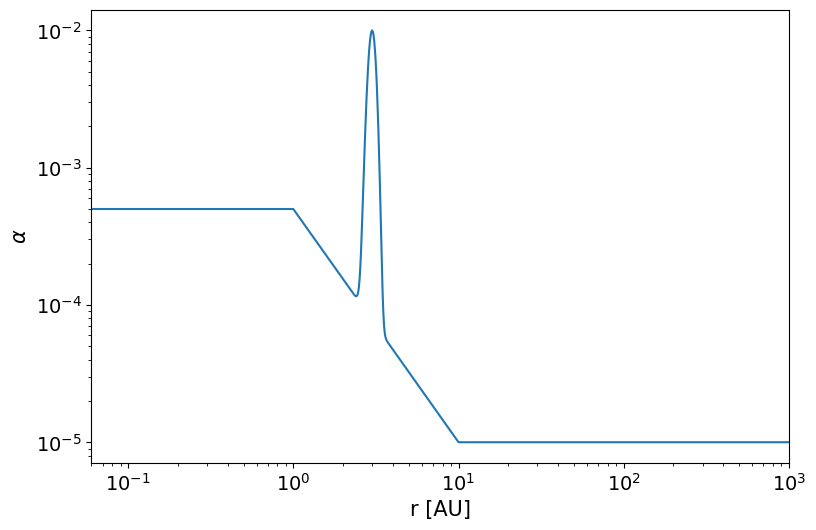}
\caption{Shakura-Sunyaev $\alpha$-parameter as a function of radius in the disk. Two regions of constant $\alpha$ in the inner and outer disk are connected by a decreasing powerlaw between 1 and 10 AU. This profile applies to the post-infall phase only. The sharp Gaussian peak at 3 AU forms at 0.6 Myr, and is caused by the presence of Jupiter's growing planetary core. This peak is not used for turbulent mixing of dust species, as it represents a torque exerted on the gas by the planet, and not a real increase in viscosity.}\label{fig:AlphaProfile}
\end{figure}

Some time after infall had ended and the disk had fully formed, Jupiter's core was assumed to form, start growing and open a gap in the disk. We once again followed the DKA18 model in the way this was incorporated into the simulation. Jupiter's core first appeared in the model when it had reached a mass of $M_J = 30$ $M_{\oplus}$. It then started growing to its full size by accreting gas from its surroundings. At every time step $dt$, an amount of mass was added that is calculated as \footnote{This quantity of gas was not actually removed from the disk, however.}

$$
dM = \frac{dt}{\tau} \int_r \Sigma_g(r) e^{-x^2} 2 \pi r dr, \eqno{(39)}
$$

\noindent where

$$
x = \frac{r-r_J}{R_H}, \eqno{(40)}
$$

\noindent with the Hill Radius $R_H$ given by

$$
R_H = r_J \left(\frac{M_J}{3 M_*}\right)^{1/3}. \eqno{(41)}
$$

\noindent Here $r_J$ is the location where the planet forms, assumed to be at 3.0 AU (or roughly 40\% closer in than its current position at 5.2 AU from the Sun) and $\tau$ is the growth time scale which sets what fraction of the available gas in the vicinity is accreted. Because the gas surface density $\Sigma_g$ is a function of radius and time and is also modified by the presence of the planet itself, the precise value of $\tau$ used depends on the choice of time step as well as $r_J$. The value we used, $\tau = 1537.5$ yr, thus only works well for our chosen time step of 1000 years for the post-infall phase. No mass was added to the planet beyond $M_J = 317.8$ $M_{\oplus}$, which equals 1 Jupiter mass. For the chosen values for $r_J$ and $\tau$, this value was reached after roughly 4.5 Myr. The way a gap was opened in the disk by this growing protoplanet is by modifying the value for $\alpha$ in its vicinity:

$$
\alpha_\text{new} = \alpha + (\alpha_\text{peak} - \alpha) e^{-x^2}. \eqno{(42)}
$$

\noindent This adds a Gaussian spike to the $\alpha$-profile as described before, which can be seen in Figure \ref{fig:AlphaProfile}. The value of $\alpha_\text{peak}$ was set to 0.01, in accordance with Desch et al (2018). This peak in $\alpha$ acts as a torque by the planet, pushing material away and out of the gap. Because physically there is no real increase in the turbulent viscosity, the mixing of dust species should not be affected. Therefore the $\alpha$-profile used in the mixing calculation does not include this peak. The final parameter relevant to the planet gap is the formation time $t_\text{planet}$, which is also the time when the gap starts to open. As in the DKA18 model, this time was set to $t_\text{planet} = 0.6$ Myr, though it must be noted that this time can't be directly compared between the two models. In the DKA18 model, $t = 0$ refers to the point where the disk is fully formed (the end of their non-existent infall phase) and the time at which CAI formation starts. In our model however, $t = 0$ corresponds to the very beginning of the infall phase. As we will see in the Results section, CAI formation already begins early in the infall phase, so well before the disk is finished building up. So while $t_\text{planet}$ has been chosen to (at least roughly) match the 0.6 Myr after the first CAIs start to appear, it cannot simultaneously match the 0.6 Myr after the end of the infall phase. Both the planet formation time $t_\text{planet}$ and the formation location $r_J$ will be varied later to see how different choices for these parameters impact the results.

At the end of the 5 Myr simulation, the disk was simply left as is. No physics was included to simulate the eventual dissipation of the disk, as this would occur after the phenomenon of interest, CAI parent body formation, has already occurred. A summary of physical parameter choices made for the main model is shown as Table 2.

\section{Results}\label{Results}

\subsection{Main model}\label{Main Model}

\begin{figure*}[t!]
\minipage[t]{\columnwidth}
  \includegraphics[width=\columnwidth]{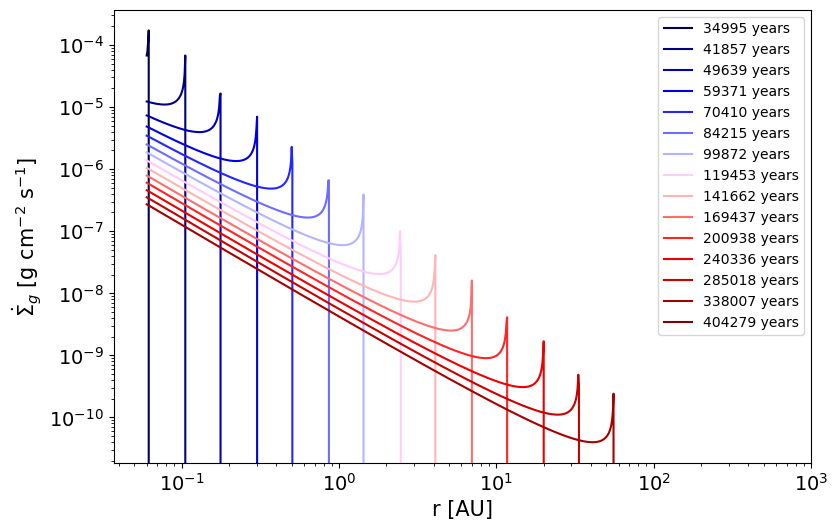}
  \caption{Infall rate $\dot{\Sigma}_g$ of gas from the molecular cloud core onto the disk. As time passes, the centrifugal radius $r_c$ increases and infalling matter is added to greater and greater radii in the disk, while the total accretion rate $\dot{M}$ remains constant. At the end of infall, the centrifugal radius has reached 93.9 AU.}\label{fig:Sigdot}
\endminipage\hfill
\minipage[t]{\columnwidth}
  \includegraphics[width=\columnwidth]{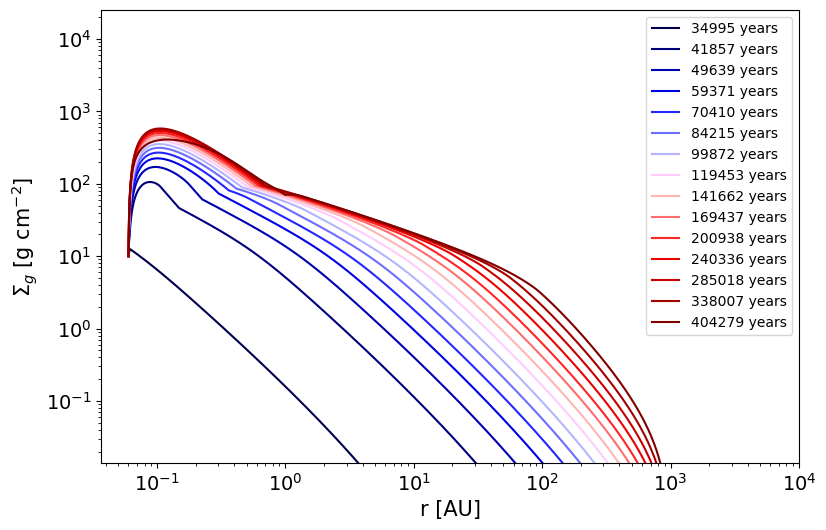}
  \caption{Viscous evolution of the gas surface density $\Sigma_g$ during the infall phase. The disk starts building up when $r_c > r_\text{in}$. Rapid viscous expansion causes the gas to move all the way out to 1000 AU. The surface density keeps increasing everywhere during the entire infall phase as more and more gas is added to the disk.}\label{fig:SigmaGasInfall}
\endminipage\hfill
\end{figure*}

\begin{figure*}[h]
\minipage[t]{\columnwidth}
  \includegraphics[width=\columnwidth]{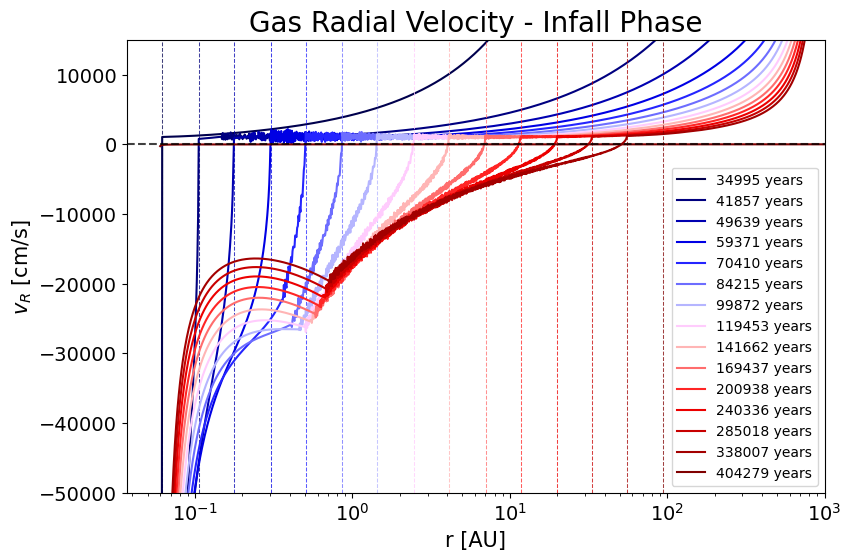}
  \caption{Radial velocity of the gas $v_R$ during the infall phase. At the centrifugal radius, indicated by the dashed vertical lines, $v_R$ becomes strongly positive due to a large gradient in surface density pushing gas outward. Interior to this radius, the gas moves inward as it accretes onto the Sun. The sudden vertical jumps are likely numerical artefacts that are not expected to have a significant impact on the results.}\label{fig:VrGasInfall}
\endminipage\hfill
\minipage[t]{\columnwidth}
  \includegraphics[width=\columnwidth]{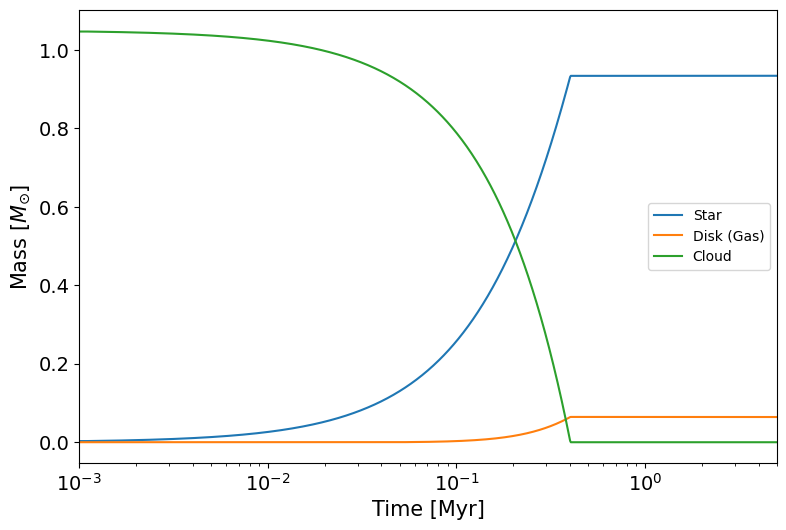}
  \caption{Mass evolution of the molecular cloud core, the Sun and the disk. The infall rate of matter on the disk is constant over time. Most infalling material quickly accretes onto the Sun, which reaches 0.93 $M_{\odot}$ at the end of the infall phase after 0.4 Myr. The solar and disk mass change little post-infall, as most of the disk mass is located at large radii where the viscosity is small.}\label{fig:MassEvolution}
\endminipage\hfill
\end{figure*}

We can now move on to the results of our main model. The disk starts building up when the centrifugal radius $r_c$ exceeds the inner edge of the disk $r_\text{in}$, which happens around 35 kyr after the start of the simulation. The infalling gas is then spread out over the disk according to Equation (32), the result of which is shown as Figure \ref{fig:Sigdot}. The centrifugal radius itself can be seen in this plot as the location of the steep vertical line beyond which no material rains down on the disk at that time. As this radius increases over time, so does the total surface area of the disk within that radius. The value of $\dot{\Sigma}$ at any one radius $r < r_c$ must therefore decrease over time, in order for the total accretion rate $\dot{M}$ to remain constant as per Equation (30). The infall phase ends when the molecular cloud core has been depleted of mass after roughly 0.4 Myr, at which point the centrifugal radius has reached $r_c = 93.9$ AU.

Figure \ref{fig:SigmaGasInfall} shows the resulting surface density evolution of the gas during the infall phase. What stands out in this plot is that at every time step, a significant amount of gas is present in the region beyond the centrifugal radius. This means that this material must have ended up there by viscous spreading. Figure \ref{fig:VrGasInfall} shows the radial velocity of the gas throughout the disk. The vertical dashed lines indicate the location of the centrifugal radius at each time. The radial velocity interior to $r_c$ is strongly negative throughout the infall phase, since most of the gas infalling from the molecular cloud core is rapidly accreting onto the growing Sun. At the centrifugal radius however, the radial velocity switches sign as the disk is expanding outward beyond this point. 

At the end of the infall phase, the disk has reached a total mass of $M_\text{disk} = 0.064$  $M_{\odot}$. While this is comparable to the disk in the DKA18 model, which has $M_\text{disk} = 0.089$  $M_{\odot}$, this mass is spread out in a very different way. Our disk extends all the way out to 1000 AU, while the disk in the DKA18 model is much more compact, its gas surface density sharply decreasing past 10 AU. In turn, the surface density in the inner part of our disk is three orders of magnitude lower than in the DKA18 model. This has consequences for the accretion rate of the disk onto the Sun: while the disk in the DKA18 model loses about half its mass in just 0.1 Myr of subsequent evolution, Figure \ref{fig:MassEvolution} shows that in our case, both the stellar and the disk mass barely change anymore after the end of the infall phase. This could simply mean that the chosen $\alpha$-parametrization is unrealistic, as the resulting viscosity is too weak to move significant quantities of mass back in towards the Sun. 

\begin{figure*}[t!]
\minipage[t]{\columnwidth}
  \includegraphics[width=\columnwidth]{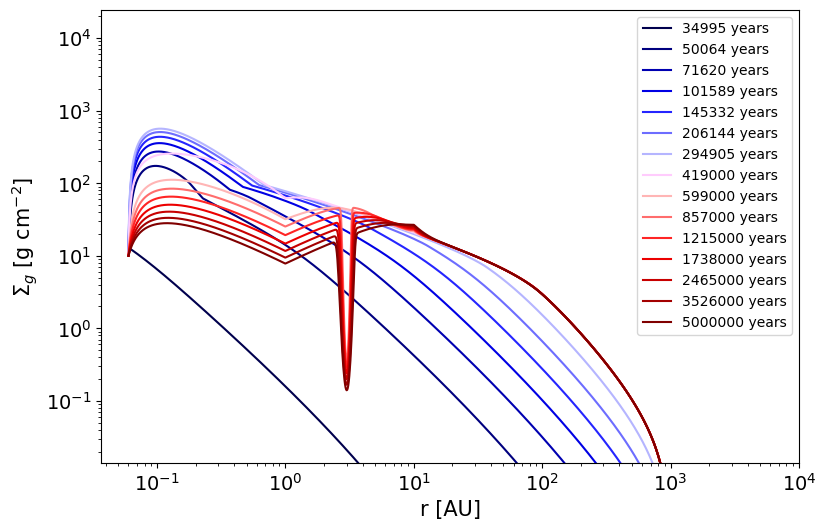}
  \caption{Viscous evolution of the gas surface density $\Sigma_g$ during the full simulation. After the infall phase, $\Sigma_g$ keeps dropping in the inner disk where matter is accreting onto the Sun. In the outer disk, the viscosity is lower and there is less observable change. The planet gap can be clearly be seen at 3 AU.}\label{fig:SigmaGas}
\endminipage\hfill
\minipage[t]{\columnwidth}
  \includegraphics[width=\columnwidth]{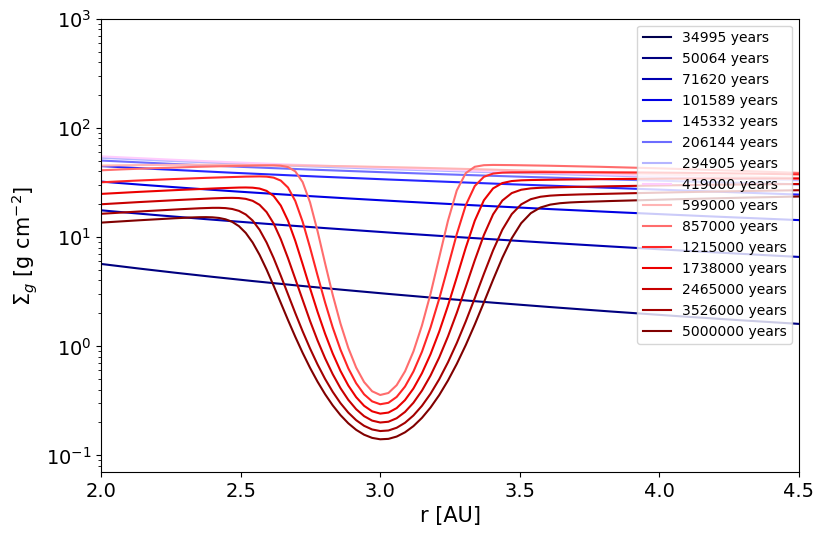}
  \caption{Close-up of Figure \ref{fig:SigmaGas} around the location of the planet gap. Surface density builds up here until the end of the infall phase, after which it starts dropping again. As Jupiter grows to its full size by accreting mass from its surroundings, its Hill radius increases, widening the gap.}\label{fig:SigmaGasPlanetGap}
\endminipage\hfill
\end{figure*}

\begin{figure*}[h]
\minipage[t]{\columnwidth}
  \includegraphics[width=\columnwidth]{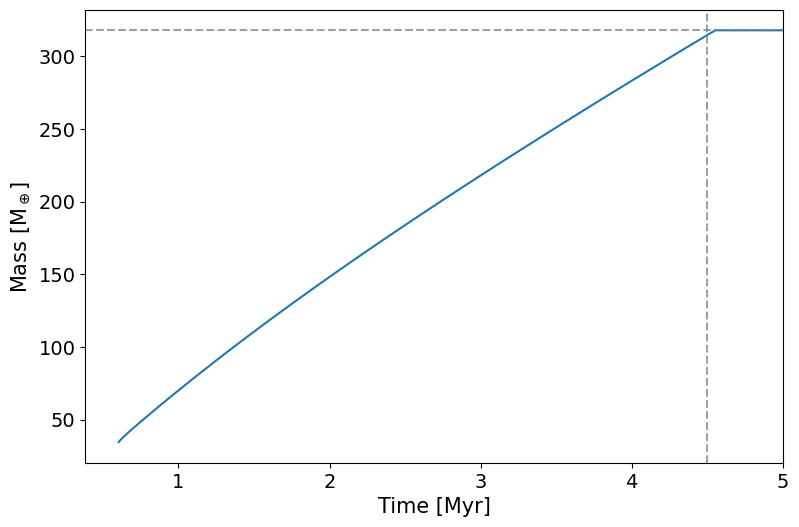}
  \caption{Growth of Jupiter from a 30 $M_{\oplus}$ core at 0.6 Myr to its full size of 317.8 $M_{\oplus}$ (1 Jupiter mass) after 4.5 Myr by accretion of gas in its vicinity. The growth turns out to be roughly linear.}\label{fig:JupiterGrowth}
\endminipage\hfill
\minipage[t]{\columnwidth}
  \includegraphics[width=\columnwidth]{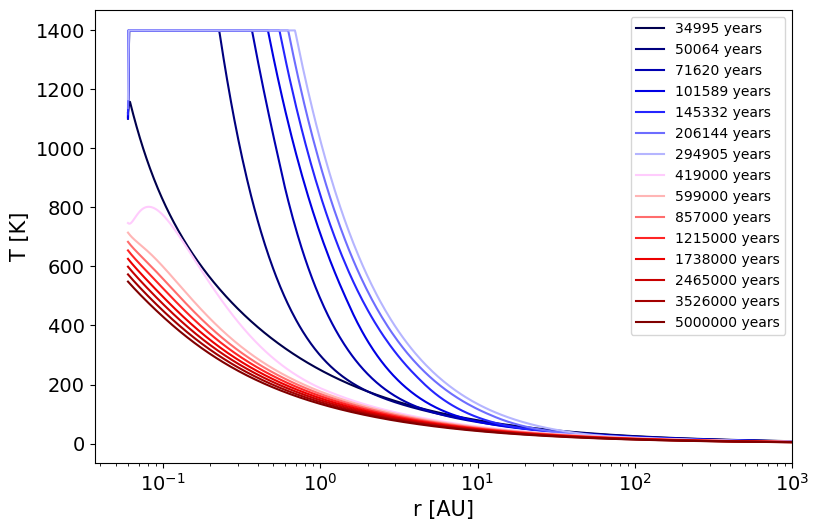}
  \caption{Midplane temperature due to irradiation and viscous heating. During the infall phase, the temperature reaches 1400 K in the inner disk, activating the CAI Factory. At the end of infall, the temperature quickly decreases again.}\label{fig:MidplaneTemperature}
\endminipage\hfill
\end{figure*}

Now that the disk has been fully formed, we can see how it evolves in the post-infall phase. Figure \ref{fig:SigmaGas} shows the full time evolution of the gas surface density from the start of disk formation to the end of the simulation. The post-infall phase is represented here by the red lines. While the surface density is clearly decreasing over time in the innermost 10 AU of the disk, little change is visible beyond that point, where $\alpha$ is lowest. An important feature of the post-infall phase is the planet gap that has opened up at $r = 3$ AU 0.6 Myr after the start of the simulation, a closeup of which can be seen as Figure \ref{fig:SigmaGasPlanetGap}. The gap can be seen to get wider over time, as Jupiter continues to accumulate mass, increasing its Hill radius. The surface density of gas within the gap is successfully reduced by 2 orders of magnitude. The mass evolution of Jupiter itself is shown as Figure \ref{fig:JupiterGrowth}. Its growth turns out to be fairly linear, with the growth time scale $\tau$ chosen so that it reaches its full mass of $M_J = 317.8$ $M_{\oplus}$ after about 4.5 Myr.

So far we have only looked at the gas component of the disk during the simulation, but what we're really interested in is the behaviour of the dust species during this time. Because this depends heavily on where and when the CAI Factory is active, we'll first look at the midplane temperature, which is shown in Figure \ref{fig:MidplaneTemperature}. The temperature in the inner disk shoots up to 1400 K soon after the disk starts building up, when viscous heating dominates the temperature calculation there. For the entire duration of the infall phase, there is then some region in the disk where the temperature reaches 1400 K, the CAI Factory. This region never extends past 1 AU, which means that CAIs can only end up past Jupiter's location of formation at $r = 3$ AU by transport within the disk, since they will not be created that far out. After the end of infall at 0.4 Myr, the temperature rapidly drops, and the CAI Factory switches off. During the post-infall phase, viscous heating is negligible compared to irradiative heating. An important result here is that the period during which CAIs are being created in the disk basically coincides with the infall phase. CAI formation should therefore have ceased by the time the disk is fully formed, unlike in the DKA18 model, where this is instead the time when the CAI Factory is first turned on. This earlier formation of CAIs significantly affects the evolution of their surface density. 

\begin{figure*}[t!]
\minipage[t]{\columnwidth}
  \includegraphics[width=0.98\columnwidth]{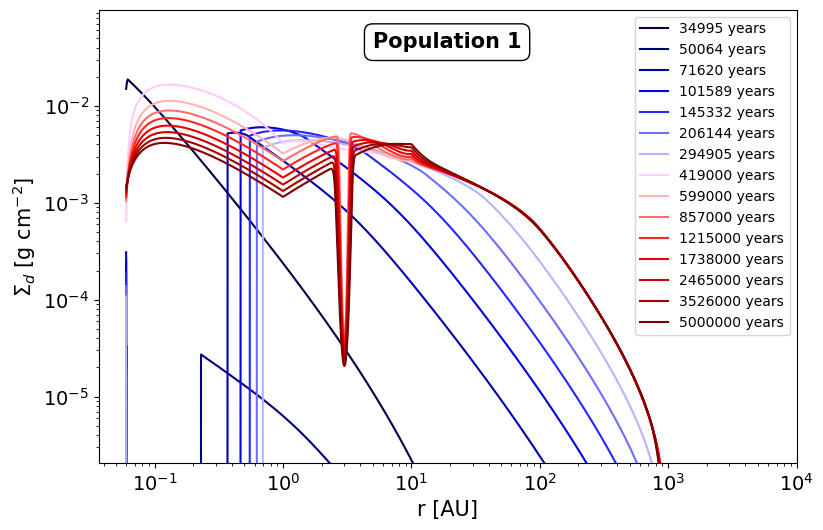}
  \caption{Surface density evolution of population 1 dust. During most of the infall phase, there is no population 1 dust in the inner disk due to its complete thermal conversion into other dust species.}\label{fig:SigmaPop1}
\endminipage\hfill
\minipage[t]{\columnwidth}
  \includegraphics[width=0.98\columnwidth]{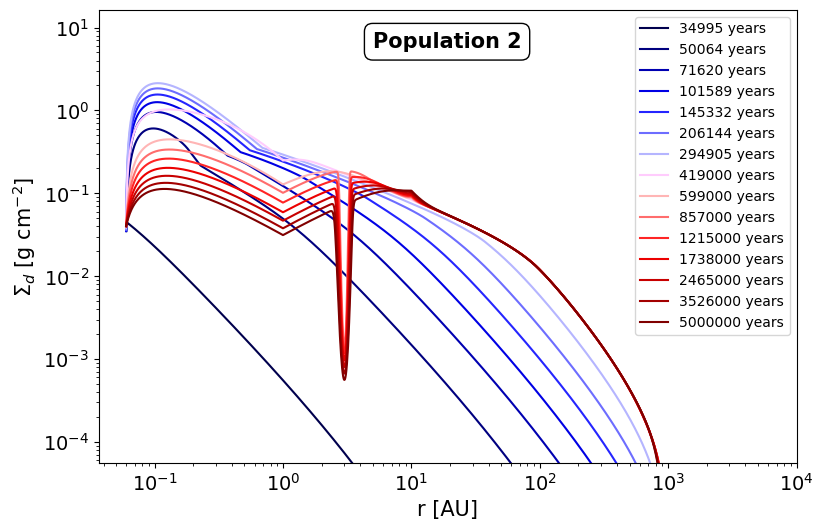}
  \caption{Surface density evolution of population 2 dust. This is the only dust species that does not take part in thermal processing in the CAI Factory.}\label{fig:SigmaPop2}
\endminipage\hfill
\end{figure*}

\begin{figure*}[h!]
\minipage[t]{\columnwidth}
  \includegraphics[width=0.98\columnwidth]{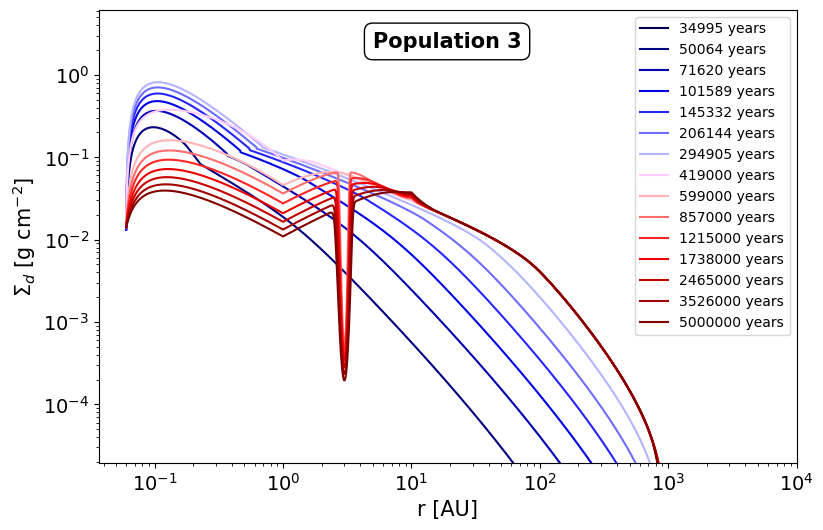}
  \caption{Surface density evolution of population 3 dust. The 35000 year line is not visible, since the temperature is still below 1400 K.}\label{fig:SigmaPop3}
\endminipage\hfill
\minipage[t]{\columnwidth}
  \includegraphics[width=0.98\columnwidth]{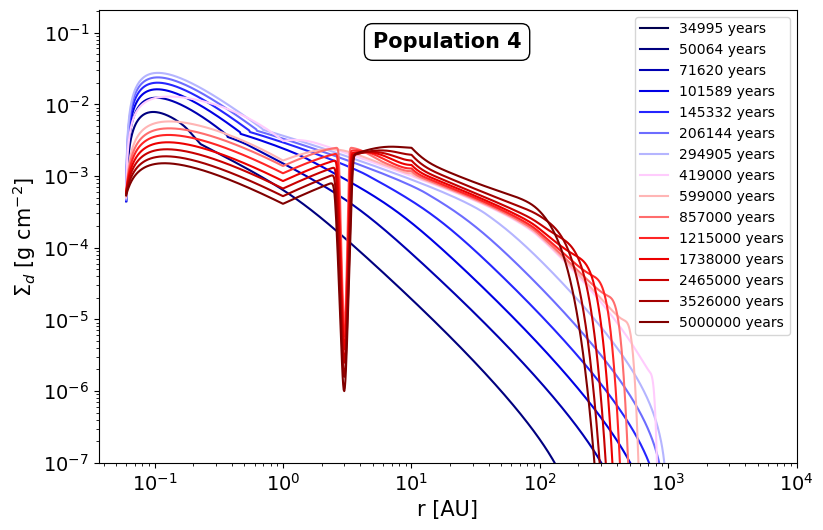}
  \caption{Surface density evolution of population 4 dust. These grains are large enough to display noticeable inward radial drift in the outer disk as well as a trapping effect in Jupiter's pressure bump. The 35000 year line is not visible, since the temperature is still below 1400 K.}\label{fig:SigmaPop4}
\endminipage\hfill
\end{figure*}

\begin{figure*}[h!]
\minipage[t]{\columnwidth}
  \includegraphics[width=0.98\columnwidth]{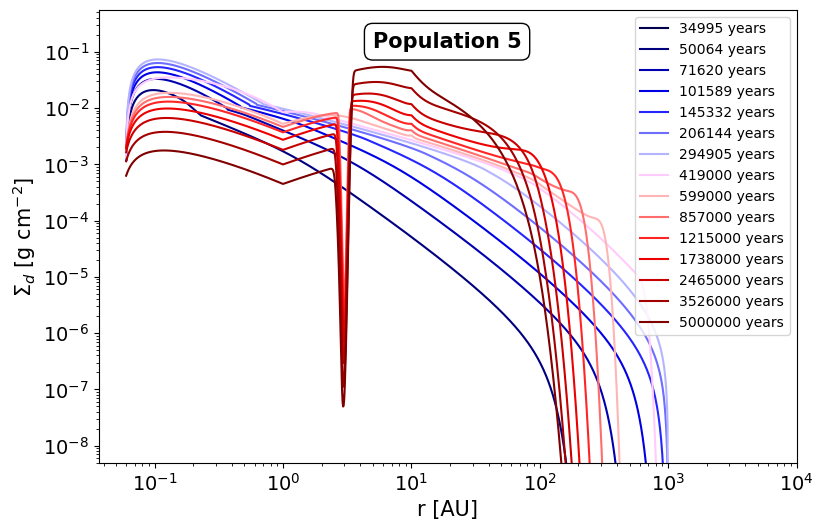}
  \caption{Surface density evolution of population 5 dust (CAIs). As all other dust species, the CAIs are efficiently transported into the outer disk during the infall phase. They then start drifting back in and piling up in the region behind the planet gap. CAIs in the inner disk slowly drain as they accrete onto the Sun. The 35000 year line is not visible, since the temperature is still below 1400 K.}\label{fig:SigmaPop5}
\endminipage\hfill
\minipage[t]{\columnwidth}
  \includegraphics[width=0.98\columnwidth]{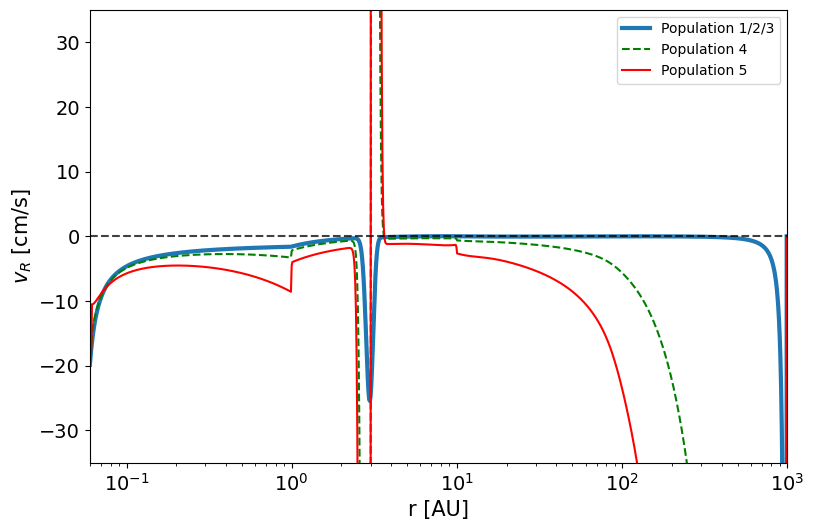}
  \caption{Radial velocity of the dust species in the post-infall phase. The micrometer-sized particles, populations 1-3, are well coupled to the gas and experience very little radial drift in the outer disk. They are not caught in the pressure bump, but simply drift through the gap into the inner disk. The larger populations 4 and 5 have higher radial drift velocities, but cannot easily pass the planet gap.}\label{fig:VrDustPostInfall}
\endminipage\hfill
\end{figure*}

\begin{figure*}[t!]
\minipage[t]{\columnwidth}
  \includegraphics[width=\columnwidth]{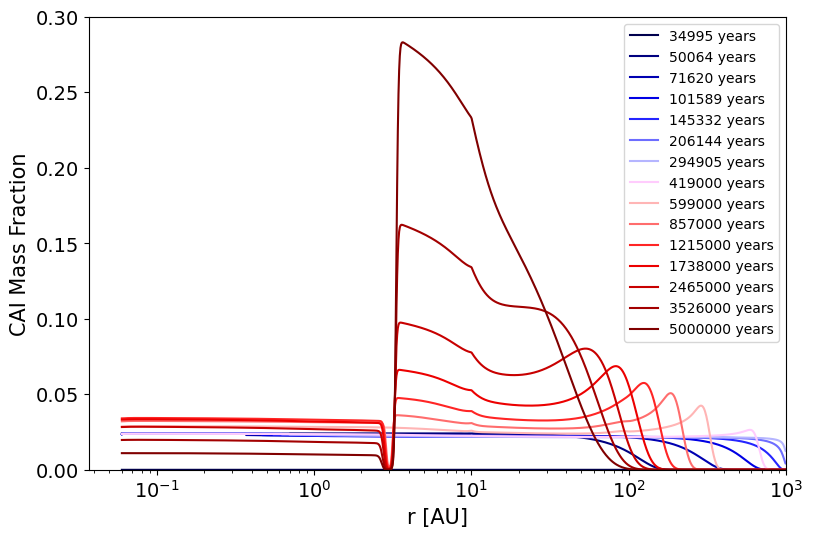}
  \caption{Time evolution of the CAI abundance throughout the disk. CAIs are created in the innermost disk and efficiently transported outward, where they are diluted with new infalling population 1 dust. As they start drifting inward in the post-infall phase, a double peak profile develops. CAIs are piling up in Jupiter's pressure bump as well as in the far outer disk, when they are slowed down by the increasing gas density. The two peaks eventually merge in the pressure bump.}\label{fig:CAIAbundance}
\endminipage\hfill
\minipage[t]{\columnwidth}
  \includegraphics[width=\columnwidth]{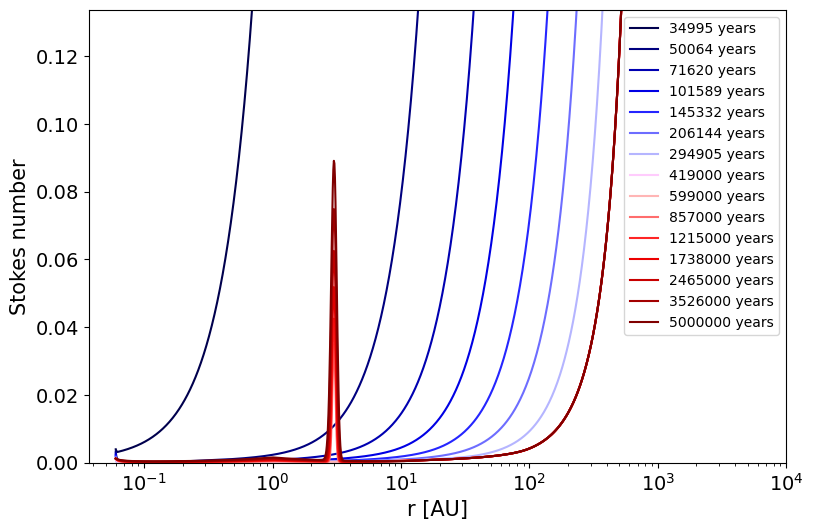}
  \caption{Stokes number of CAIs, a dimensionless coupling constant that indicates how well the dust grains are coupled to the gas. Lower values of the Stokes number imply better coupling to the gas and a shorter stopping time $\tau_\text{stop}$, which evens out velocity differences between the dust and gas. For a Stokes number that is zero, the radial drift velocity of the gas $v_d$ equals the radial velocity of the gas itself $v_r$, as shown by Equation (25).}\label{fig:CAIStokes}
\endminipage\hfill
\end{figure*}

Figures \ref{fig:SigmaPop1}, \ref{fig:SigmaPop2}, \ref{fig:SigmaPop3}, \ref{fig:SigmaPop4}, and \ref{fig:SigmaPop5} show the surface density evolution for the five different dust species in the model. As Figure \ref{fig:SigmaPop1} shows, population 1 dust is completely absent in the innermost part of the disk during the infall phase (except for the very first line when the temperature has not quite reached 1400 K) as that is where it is being converted into populations 3, 4, and 5. An important thing to note is that, just as with the gas, each of the different populations is present much further out in the disk than the centrifugal radius. Evidently the dust particles are being dragged along with the gas during the early rapid viscous expansion phase as well as experiencing strong mixing. This is true even for populations 3, 4, and 5 (CAIs), which originate exclusively from the CAI Factory in the innermost part of the disk. At the end of the infall phase then, each dust population can be found all the way out to 1000 AU. From this point on, the dust particles start to drift back in towards the Sun. Because populations 1, 2, and 3 are micron-sized particles that are well coupled to the gas, their radial velocity is essentially equal to that of the gas throughout the disk. Similarly to the gas in Figure \ref{fig:SigmaGas} then, these dust species slowly drain from the inner disk, but little change in their surface density can be seen beyond 10 AU. 

\textcolor{white}{--- --- --- --- --- --- --- --- --- --- --- --- --- --- --- --- --- --- --- --- --- --- --- --- --- --- --- --- --- --- --- --- --- --- --- --- --- --- --- --- --- --- --- --- --- --- --- --- --- --- --- --- --- --- --- --- --- --- --- --- --- --- --- --- --- --- --- --- --- --- ---}

\noindent No trapping effect can be seen for these populations behind the planet gap. The sign of the pressure-gradient force reverses in the outer gap, where the gas pressure increases in the outward direction, which increases the inward force on the gas and causes it to orbit with super-Keplerian velocities. The resulting drag force on solid particles causes them to drift towards the pressure maximum, away from the star. However, smaller particles such as populations 1 through 3 are coupled to the gas strongly enough to be dragged along with it as it flows through the gap, while only larger particle sizes with radial velocities exceeding that of the gas are hindered by this barrier. This process is called dust filtration by Rice et al (2006). In Figure \ref{fig:VrDustPostInfall}, which shows the effective radial velocity of the different dust species in the post-infall phase, it is shown that dust particles belonging to populations 1, 2, and 3, when approaching the planet gap from larger radii, are simply accelerated through. The picture is different for the much larger AOAs (population 4) and CAIs (population 5). These populations have higher Stokes numbers than small dust grains, and as such they drift more strongly towards higher gas pressures, leading to larger radial drift velocities. In the far outer disk, the radius at which the surface density of these dust species starts to drop off steeply can be seen to move inward over time. Because the planet gap does serve as an effective barrier against particles of these sizes, they are piling up in the region just behind the planet gap, as evidenced by the surface density increasing over time there. Similar to the other dust populations, the CAIs and AOAs are slowly depleting in the inner disk, where no barrier exists to prevent them from drifting into the Sun. However, the average radial velocity of CAIs in the inner disk implies that all CAIs would vanish in this region on a time scale of order 10$^5$ years after their formation ceased. Since this is clearly not what is happening, with a lower but still significant amount of surface density left even after 5 Myr, at least some part of the CAIs must still be leaking through the planet gap into the inner disk.

Figure \ref{fig:CAIAbundance} shows the CAI abundance as a mass fraction of all the dust species in the disk. This represents the main result of this project. Comparing first the final abundance profile after 5 Myr to the result from the DKA18 simulation (their Figure 8), we see that they are roughly similarly shaped, with a large abundance peak just beyond Jupiter's location and little to no CAI abundance elsewhere. There are however also some important differences. First of all, the peak in our model is much broader, extending over almost 100 AU, while the peak in the DKA18 model is not even one full AU wide. This can at least be partially explained by the CAIs having been transported outward so far that they are simply still in the process of drifting back in. Given more time (or equivalently, given a higher accretion rate in the outer disk), the CAIs would presumably continue to drift back in, leading to a narrower and taller abundance peak. Second, the overall abundance in our peak is significantly higher than in the DKA18 model. However, there are other uncertainties affecting the precise abundance values. One of these is parameter choice, in particular the molecular cloud parameters. We will explore the consequences of different parameter choices in section \ref{Parameter Search}. For now, suffice it to say that the general shape of the abundance profile is more reliable than the precise quantitative values.

\begin{figure*}[t!]
\minipage[t]{\columnwidth}
  \includegraphics[width=\columnwidth]{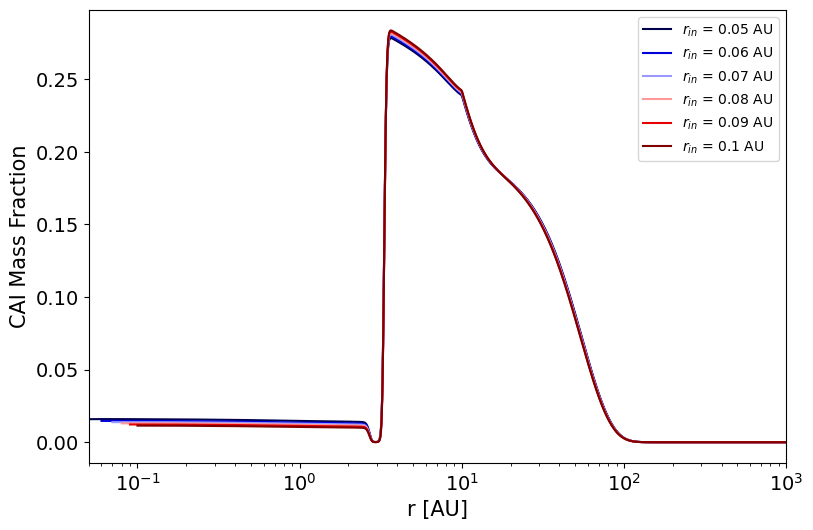}
  \caption{Abundances after 5 Myr for different values of the disk inner edge $r_\text{in}$. Higher values of $r_\text{in}$ slightly increase the peak CAI abundance, but this effect is so small that the results are basically independent of the inner edge location.}\label{fig:ParameterSearch0}
\endminipage\hfill
\minipage[t]{\columnwidth}
  \includegraphics[width=\columnwidth]{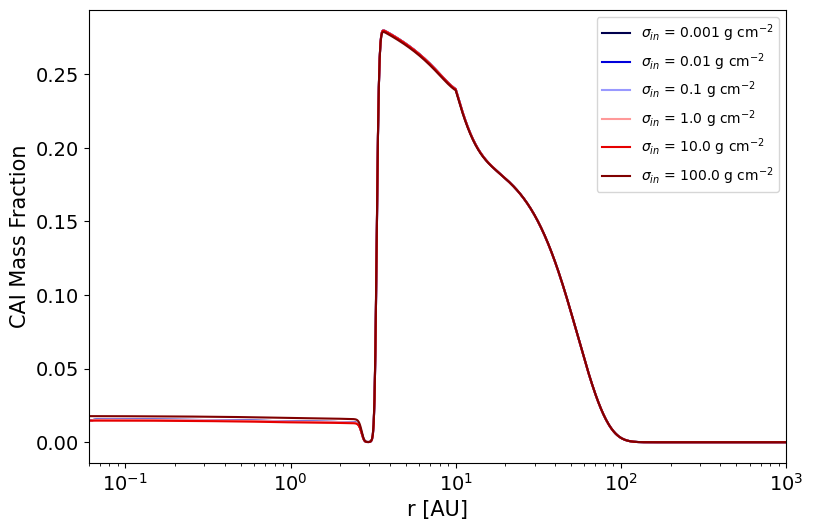}
  \caption{Abundances after 5 Myr for different values of the surface density at the disk inner edge $\sigma_\text{in}$. This parameter seems to have no impact on the CAI abundance, as it affects dust species equally.}\label{fig:ParameterSearch1}
\endminipage\hfill
\end{figure*}

\begin{figure*}[h!]
\minipage[t]{\columnwidth}
  \includegraphics[width=\columnwidth]{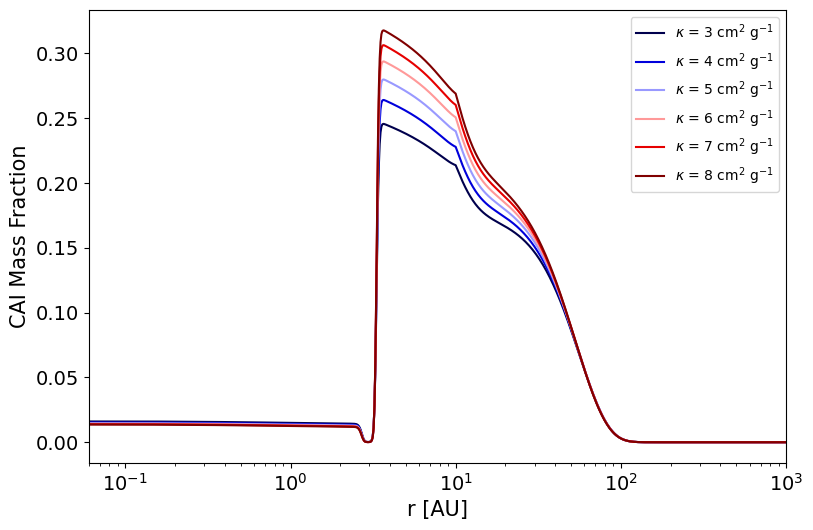}
  \caption{Abundances after 5 Myr for different values of the disk opacity $\kappa$. Higher opacities trap more heat in the disk, increasing the temperature and thereby enlarging the radial extent of the CAI Factory. This creates more CAIs and increases their abundance in the pressure bump.}\label{fig:ParameterSearch3}
\endminipage\hfill
\minipage[t]{\columnwidth}
  \includegraphics[width=\columnwidth]{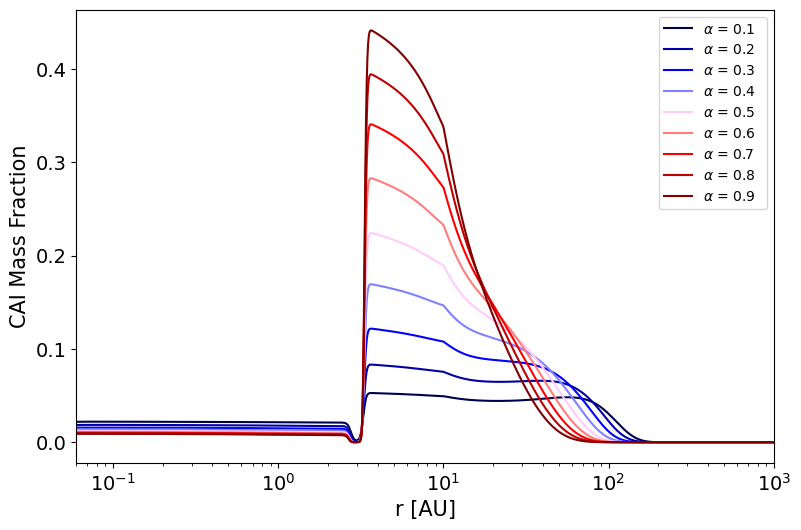}
  \caption{Abundances after 5 Myr for different values of the $\alpha$-parameter during the infall phase. Higher values increase the peak abundance, likely due to more efficient outward mass redistribution.}\label{fig:ParameterSearch8}
\endminipage\hfill
\end{figure*}

While the final abundance profile after 5 Myr looks broadly similar to that in the DKA18 model, the intermediate time steps do not. At $t = 50$ kyr, the first line in Figure \ref{fig:CAIAbundance} with a non-zero abundance, we see that the CAI abundance has a virtually constant value of 2.4\% out to 50-60 AU. Since this is the value obtained inside the CAI Factory when all population 1 dust is thermally processed, it means that outward mixing is very rapid and efficient. At subsequent time steps, we see that the CAIs are mixed further outward, until infall ends after $t = 0.4$ Myr. At this point the abundance is rather low throughout the entire disk, but inward radial drift of CAIs now commences. Something interesting then happens that does not occur in the model without infall: apart from the abundance peak that starts to build up when the planet gap opens, a second peak forms in the far outer disk. As CAIs drift in, they start slowing down (Figure \ref{fig:VrDustPostInfall}) when the gas density increases and their Stokes number (Figure \ref{fig:CAIStokes}) drops, causing a pile-up. This second peak slowly drifts inward over time and eventually merges with the first peak, leading to the final abundance profile with a single broad peak.

\subsection{Parameter search}\label{Parameter Search}

The question now arises how the results in Figure \ref{fig:CAIAbundance} depend on the chosen parameter values for the model. To determine this, a parameter search was conducted in which the simulation was run a number of times, varying one physical parameter at a time over a number of possible values. Not every possible parameter was varied in this way. For example, while the molecular cloud temperature $T_c$ and rotation rate $\Omega_c$ are not known a priori, the cloud mass $M_c$ should roughly be the same as the total mass of the Solar System, so it makes little sense to see what happens for completely different cloud masses. 

We first show what happens when varying some of the parameters associated with the disk itself. Figure \ref{fig:ParameterSearch0} shows the CAI abundance profile resulting from variation of the disk inner edge $r_\text{in}$ between 0.05 and 0.1 AU. Moving the inner edge further away from the Sun leads to slightly higher abundances in the CAI peak, but this effect is so small that we can say the overall abundance is essentially independent from the inner edge.

Figure \ref{fig:ParameterSearch1} shows the result of variations of $\sigma_\text{in}$, the gas surface density at the inner edge, which is used as a boundary condition for Equation (13). The explanation here can be brief: while this parameter influences the surface densities in the inner disk, the dust species are all affected equally, so the abundance profile does not change.

Figure \ref{fig:ParameterSearch3} shows the result for variations of the opacity $\kappa$ in the disk. When the opacity increases, it becomes harder for heat to escape from the disk, leading to an increase in the temperature, at least in the part of the disk where the temperature wasn't already at the maximum capped value. In practice then, this means that the CAI Factory grows in radial extent. More population 1 dust is then converted into CAIs, leading to a higher peak abundance for higher values of $\kappa$.

\begin{figure*}[t!]
\minipage[t]{\columnwidth}
  \includegraphics[width=\columnwidth]{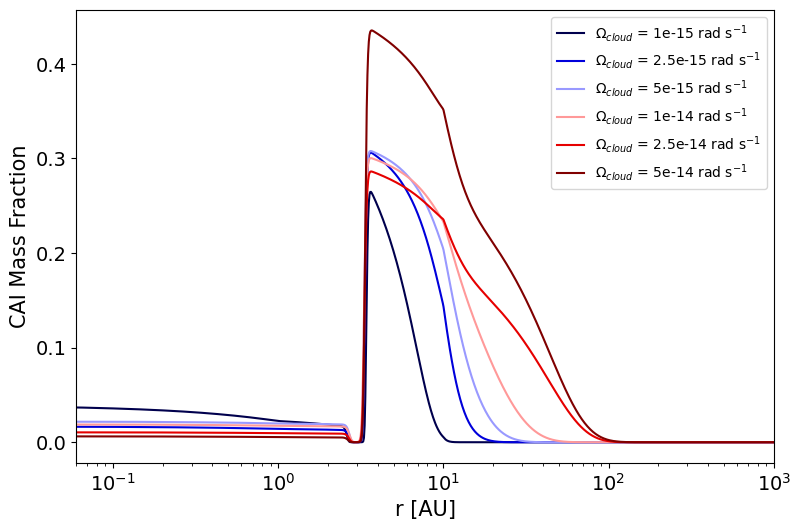}
  \caption{Abundances after 5 Myr for different values of the cloud rotation rate $\Omega_c$. The abundance peak grows taller and wider for increasing values of $\Omega_c$, as more material ends up in the disk instead of directly accreting onto the Sun, also increasing the efficiency of the CAI Factory.}\label{fig:ParameterSearch6}
\endminipage\hfill
\minipage[t]{\columnwidth}
  \includegraphics[width=\columnwidth]{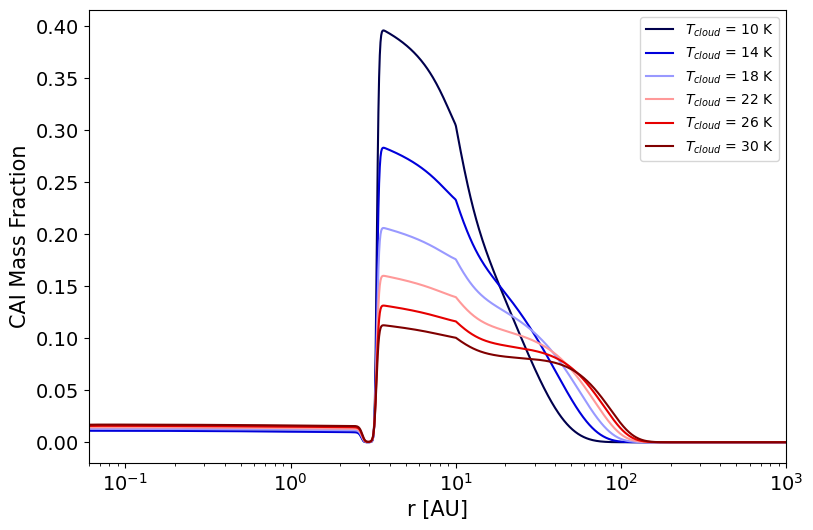}
  \caption{Abundances after 5 Myr for different values of the molecular cloud temperature $T_c$. Higher temperatures decrease the duration of the infall phase. While this causes higher surface densities in the inner disk early on, increasing the efficiency of the CAI Factory, it also causes stronger outward transport of CAIs, spreading them further out over the disk.}\label{fig:ParameterSearch5}
\endminipage\hfill
\end{figure*}

\begin{figure*}[h!]
\minipage[t]{\columnwidth}
  \includegraphics[width=\columnwidth]{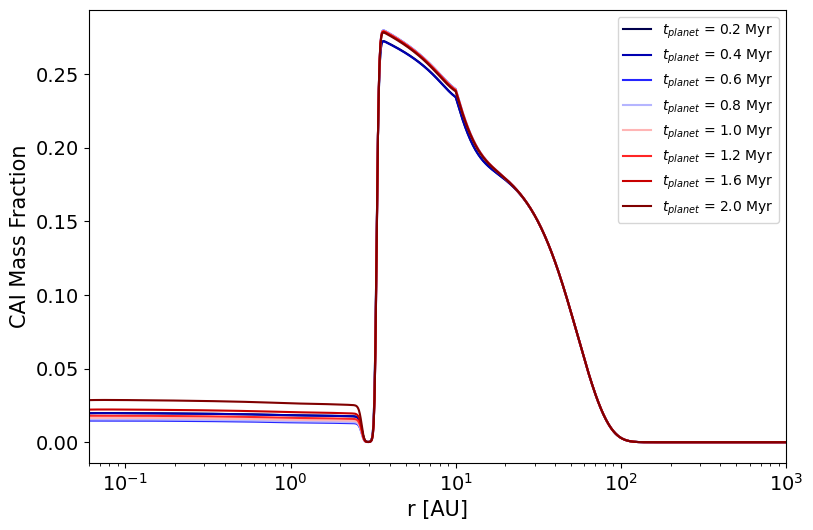}
  \caption{Abundances after 5 Myr for different values of the planet formation time $t_\text{planet}$. Perhaps surprisingly, this parameter has little effect on the final abundance profile. CAIs are mixed so far out that they are still in the process of drifting back in even after 5 Myr.}\label{fig:ParameterSearch10}
\endminipage\hfill
\minipage[t]{\columnwidth}
  \includegraphics[width=\columnwidth]{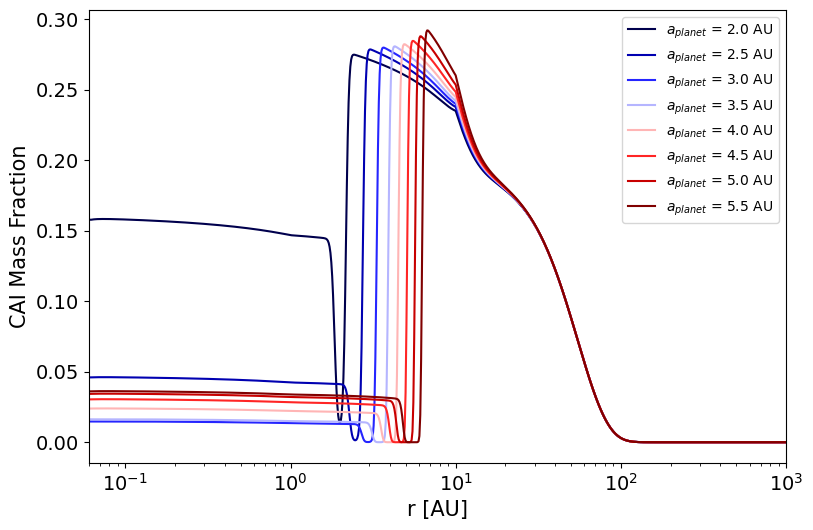}
  \caption{Abundances after 5 Myr for different values of the planet formation location $a_\text{planet}$. As the planet moves outward, its Hill radius increases, widening the gap and making the trapping slightly more efficient.}\label{fig:ParameterSearch11}
\endminipage\hfill
\end{figure*}

The search over different values of the initial $\alpha$-parameter during the infall phase is presented as Figure \ref{fig:ParameterSearch8}. A trend is visible where the peak abundance increases for higher values of this parameter. Since this $\alpha$ is used to mimic the effects of gravitational instability, higher values imply stronger redistribution of mass in the disk. This might transport additional CAIs outward, increasing abundances in the outer disk. What is less clear is why the abundance peak is broader for lower values of $\alpha$. Perhaps the lower efficiency of mass redistribution slows down the inward radial drift of CAIs in the outer disk because the gas surface density ends up lower, reducing drag forces on the dust. It must be noted that for the lowest values of $\alpha$, the disk technically becomes too massive to ignore the effects of self-gravity, even though these values do follow the general trend of lower but broader abundance peaks in this parameter search. For $\alpha = 0.1$ or 0.2, the disk mass exceeds 0.15 M$_{\odot}$. This might also have an impact on the results.

Moving on to the molecular cloud parameters, Figure \ref{fig:ParameterSearch6} shows the dependence of the CAI abundance on variation of the molecular cloud rotation rate $\Omega_c$, which is also a measure of the total angular momentum in the cloud. Here we see that the CAI abundance peak past the planet gap grows both wider and taller for increasing angular momentum. The first of these effects is easy to understand. If the angular momentum in the cloud is low, the centrifugal radius in Equation (31) is also low, and infalling material is deposited on the disk relatively close to the star. In contrast, high cloud angular momentum causes infalling material to be deposited further out. A secondary effect of this is that, because it takes longer for the centrifugal radius to reach the inner edge of the disk when $\Omega_c$ is low, more of the cloud mass will accrete directly onto the Sun instead of on the disk, leading to a less massive disk. The second effect, higher peak abundances for higher $\Omega_c$, is more surprising, because in general, rapidly rotating molecular clouds produce more massive disks with lower crystallinity than slower rotating clouds (Dullemond et al. 2006). We would expect that first, if more infalling population 1 dust ends up at larger radii from the Sun, less of it ends up in the CAI Factory where it can be used to produce CAIs. The total amount of CAIs in the disk will then also be lower. Second, population 1 dust infalling at larger radii will dilute the CAI abundance at those locations. These two effects would lead to lower abundances behind the planet gap for larger values of $\Omega_c$ instead of higher. However, it seems these effects are overshadowed by another: for higher $\Omega_c$, surface densities in the disk will be higher further out, leading to more viscous heating and higher temperatures there as well. This increases the radial extent of the CAI Factory, which produces CAIs efficiently enough that the net effect is an increase in abundance.

\begin{figure*}[t!]
\minipage[t]{\columnwidth}
  \includegraphics[width=\columnwidth]{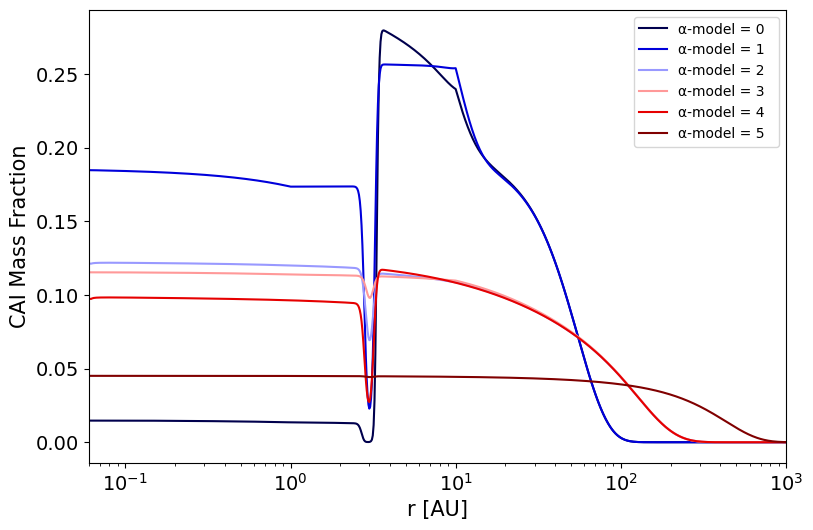}
  \caption{Abundances after 5 Myr for different values of the post-infall $\alpha$-profile. See the text for the meaning of the different $\alpha$-models.}\label{fig:ParameterSearch9}
\endminipage\hfill
\minipage[t]{\columnwidth}
  \includegraphics[width=\columnwidth]{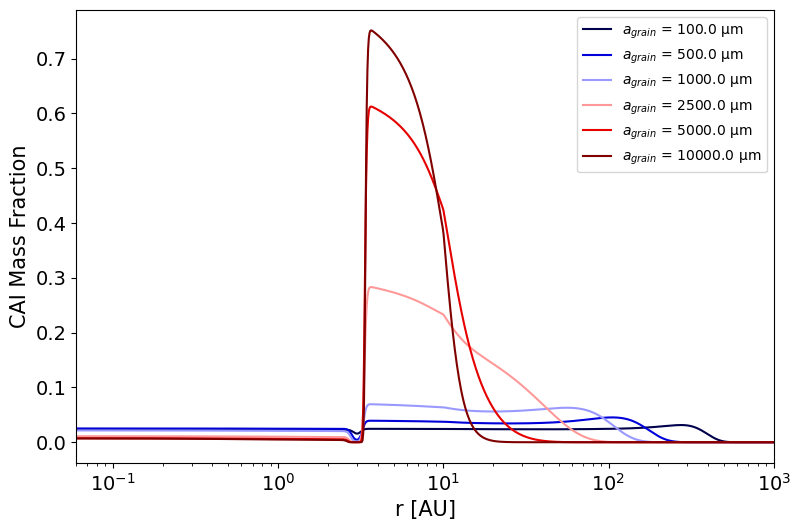}
  \caption{Abundances after 5 Myr for different values of the CAI grain size $a_\text{grain}$. The larger the grain size, the faster the CAIs drift back in to pile up in the pressure bump. Smaller grains are more evenly spread throughout the disk.}\label{fig:ParameterSearch14}
\endminipage\hfill
\end{figure*}

Figure \ref{fig:ParameterSearch5} shows the parameter search over the molecular cloud temperature $T_c$, which seems to have a larger effect on the peak CAI abundance than most other parameters searched over. Here we must first note that varying the cloud temperature alone would have an undesirable by-effect. When the cloud temperature increases, so does the local sound speed within the cloud. This means that the outward travelling expansion wave that causes the cloud to collapse moves faster, so the entire infall phase takes place in a shorter time span. However, the radius of the cloud also depends on the sound speed through

$$
R_c = \frac{G M_c}{2 c_s^2}. \eqno{(43)}
$$

\noindent Therefore, increasing the temperature of the molecular cloud will cause it to contract. However, because the rotation rate $\Omega_c$ is kept constant, this removes angular momentum from the cloud. We therefore varied both the cloud temperature and rotation rate at the same time, to keep the total angular momentum constant. We see that higher cloud temperatures then lead to a lower but broader CAI abundance peak. At least early on, the more rapid infall should lead to higher surface densities in the inner disk, making CAI production more efficient. However, the higher surface densities also lead to stronger outward transport, and this effect seems to dominate here. In the simulations with higher molecular cloud temperatures, the CAIs have been spread out more, leading to a lower but broader abundance peak which would presumably become taller and narrower given more time for continued inward radial drift. 

Moving on to the planet parameters, there are two important quantities we have varied: the planet formation time $t_\text{planet}$, shown in Figure \ref{fig:ParameterSearch10}, and the planet formation location $a_\text{planet}$, shown in Figure \ref{fig:ParameterSearch11}. Interestingly, while changing the location of formation obviously also moves the location of the planet gap and the pressure maximum beyond it, the final abundance profile otherwise does not change very strongly with these parameters. If the effects of infall would not be considered, this would be rather surprising. In the DKA18 model, CAIs are transported outward only by means of turbulent diffusion and meridional flow. Their surface density rapidly drops beyond $r = 3$ AU, which means that much less CAIs would be present beyond the location of the planet gap if Jupiter formed further out in the disk, leading to a lower peak abundance. Likewise, because CAI formation has ceased by the time the planet forms, later formation times would mean that more CAIs would have drifted back in towards the Sun, possibly already having accreted onto it. This again leaves less CAIs far enough out to be trapped in the pressure maximum. To achieve the CAI abundance profile from the DKA18 model then, it is essential that Jupiter does not form too late or too far from the Sun. In our model however, the outward transport of CAIs during the infall phase is so efficient that CAIs will remain present in the outer disk, drifting in, even after several million years. The precise time of formation (even when infall is still ongoing at 0.2 Myr) or location then make a relatively small difference to the final abundances. 

While by no means an exhaustive search over all the different possibilities, we can also attempt to run the model with different parameterizations of the post-infall $\alpha$. The result of this is shown as Figure \ref{fig:ParameterSearch9}. For reference, the default $\alpha$-profile as given by Equations (36) through (38) is shown as $\alpha$-model 0. In $\alpha$-model 1, the viscosity was raised by a factor 10 in the inner disk ($r < 1 $ AU), with the powerlaw part between 1 and 10 AU adjusted to correctly connect the two regions of constant $\alpha$. The primary effect of this model seems to be to increase CAI abundances in the inner disk, where stronger mixing likely counteracts the effects of radial drift inward. In $\alpha$-model 2, the viscosity was increased by a factor 10 in the outer region ($r > 10$ AU) instead. This seems to decrease the effectiveness of the particle trap in the pressure bump, as the stronger turbulent mixing behind the planet gap has an easier time propelling CAIs through the gap into the inner disk, where the abundance is now comparable to that in the pressure bump. This is similar to the situation in $\alpha$-model 3, where the entire $\alpha$-profile was increased by a factor 10. In $\alpha$-model 4, a constant value of $\alpha = 10^{-4}$ was used throughout the disk, while $\alpha$-model 5 represents the case with $\alpha = 10^{-3}$. Both of these cases lead to a situation where the CAI abundance interior to the planet gap is not much different from that exterior to it. None of these models therefore really represents an improvement over the $\alpha$-profile used for our main model, which most accurately reproduces the observations of high CAI abundances in carbonaceous chondrites and low abundances in meteorites originating from the inner disk.

\begin{figure*}[t!]
\minipage[t]{\columnwidth}
  \includegraphics[width=\columnwidth]{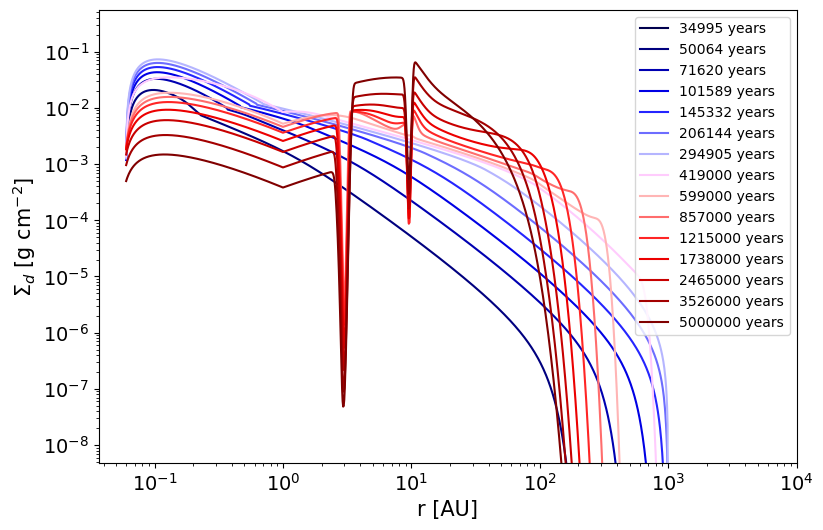}
  \caption{CAI surface density when Saturn is included into the model. While CAIs are piling up in Saturn's pressure maximum around 10 AU, some part of the CAIs is still leaking through this gap, as the surface density also keeps increasing in the region in between the two planets.}\label{fig:SaturnSigmaPop5}
\endminipage\hfill
\minipage[t]{\columnwidth}
  \includegraphics[width=\columnwidth]{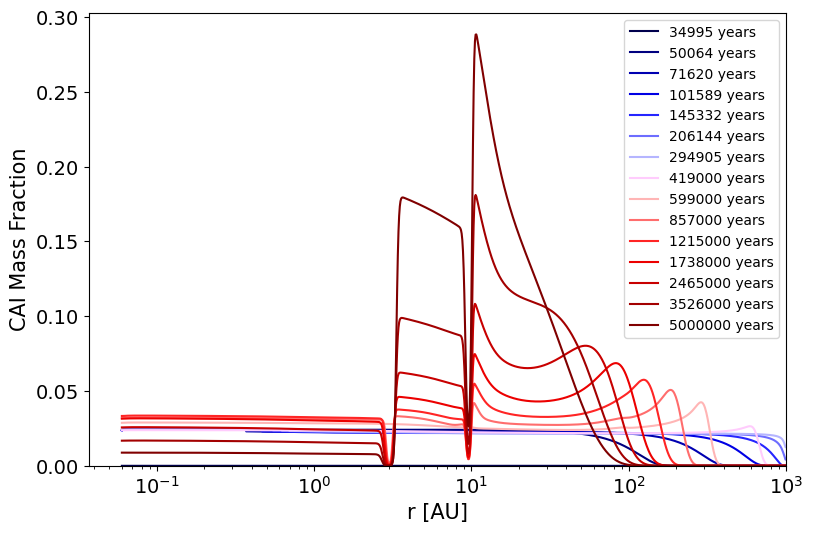}
  \caption{Evolution of the CAI abundance in the two-planet case. The largest abundance peak has shifted from Jupiter to Saturn, but enough CAIs leak through that the abundance behind Jupiter is still half of what it is in the one-planet case.}\label{fig:SaturnCAIAbundance}
\endminipage\hfill
\end{figure*}

Our main model only takes CAIs with a grain size of \SI{2500}{\micro\metre} into account. While most of the mass in CAIs is contained in such large grains, they are found almost entirely in one group of carbonaceous chondrites (CV), while the most widely occurring type of CAI consists of much smaller grains (Cuzzi et al. 2003). A final thing we can try then, is to see how our model impacts CAIs of different sizes. This is shown in Figure \ref{fig:ParameterSearch14}. CAIs up to \SI{1000}{\micro\metre} remain spread out over the outer disk quite evenly, while larger CAI grains, which have higher drift velocities, pile up at the location of the pressure bump. This effect gets stronger for larger grain sizes. Smaller CAI grains also maintain a (low but noticeable) presence in the inner disk, reproducing the observation that smaller grains occur in more meteorite types, also those originating in the inner disk.

An important conclusion we can draw from the parameter searches we have shown is that our model is not very sensitive to parameter choices. For a particular grain size, many of the parameters we have varied only have a limited effect on the CAI abundances. The molecular cloud parameters, $\Omega_c$ and $T_c$, appear to have the largest quantitative impact on the results. But importantly, the effect of particle trapping in the pressure bump is not disrupted by changes in any of the parameters we varied, possibly with the exception of different $\alpha$-parameterizations. So while quantitative uncertainties are introduced into the CAI abundances we find due to uncertainties in the parameter choices, at least qualitatively the result that the Jupiter solution keeps CAIs trapped in the pressure bump seems to be quite general. 

\subsection{A second planet}\label{A Second Planet}

The efficient outward transport of CAIs during the infall phase, and subsequent inward radial drift, raises the question how they would be affected by the presence of multiple planet gaps in the disk. This question never occurred in the case of the DKA18 model, since CAIs were just barely diffusing past the planet gap from the inner disk, instead of approaching it from the far outer disk.  If CAIs become trapped in a pressure bump caused by one of the other giant planets in our Solar System, this might prevent the build-up of a significant CAI abundance near Jupiter, which is where the carbonaceous chondrites are thought to have formed. 

\begin{figure*}[t!]
\minipage[t]{\columnwidth}
  \includegraphics[width=\columnwidth]{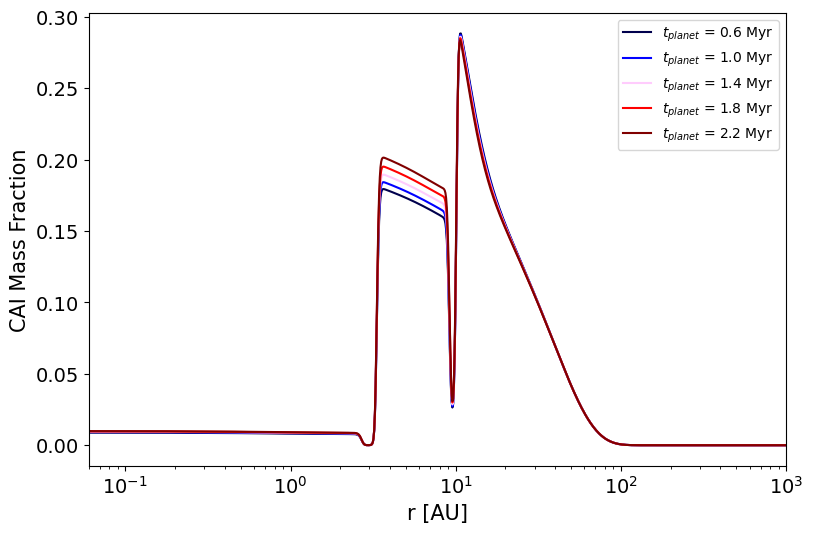}
  \caption{Final abundances when varying Saturn's formation time. Later formation allows additional CAIs to drift towards Jupiter.}\label{fig:SaturnParameterSearch0}
\endminipage\hfill
\minipage[t]{\columnwidth}
  \includegraphics[width=\columnwidth]{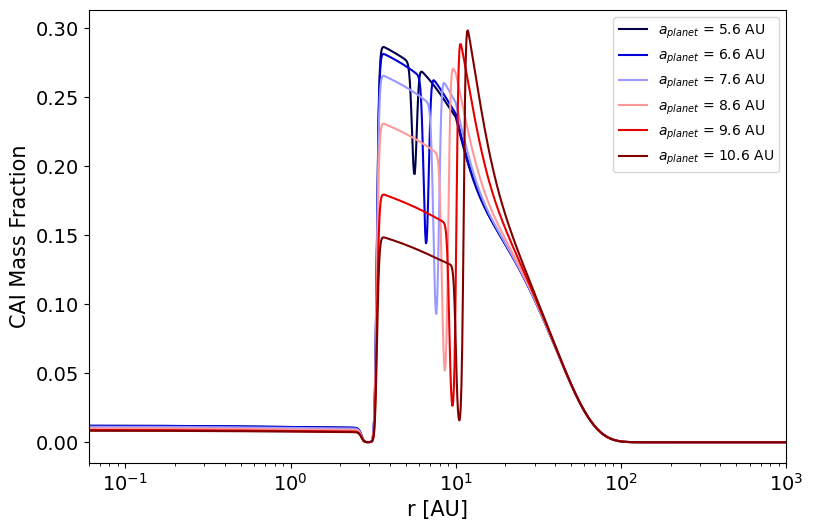}
  \caption{Final CAI abundance profile after 5 Myr with variations of Saturn's location of formation $a_\text{planet}$. The planet's Hill radius shrinks when it moves in towards smaller heliocentric distances, reducing the width of the resulting planet gap. It then becomes easier for turbulent motions in the gas to propel CAIs through the gap into the region between Jupiter and Saturn. If Saturn formed at 5.6 AU (at least for the fixed value of $\alpha_\text{peak} = 10^{-4}$), it fails in trapping many CAIs at all, and the result greatly resembles the one-planet case.}\label{fig:SaturnParameterSearch1}
\endminipage\hfill
\end{figure*}

\begin{figure}[b!]
  \centering
  \includegraphics[width=\columnwidth]{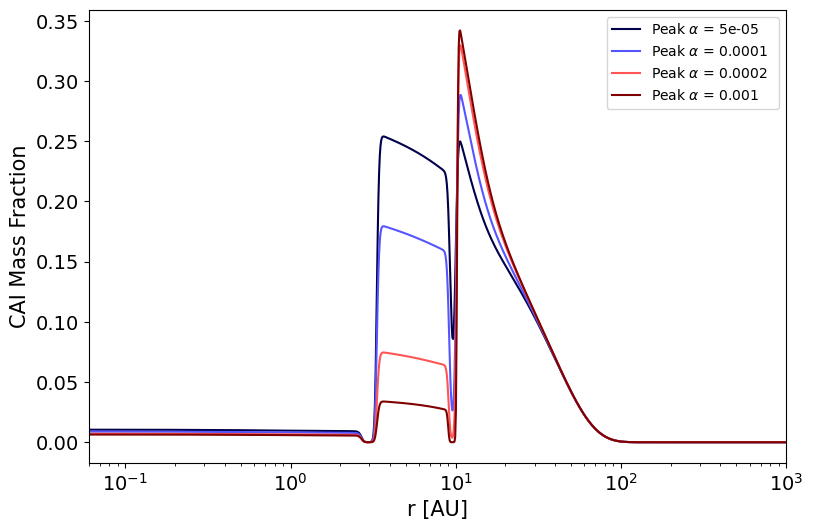}
\caption{Final CAI abundance profile after 5 Myr with variations of $\alpha_\text{peak}$, which controls the depth of the planet gap. Higher values increase the gradient in the gas surface density and hence lead to a stronger pressure gradient pushing CAIs back towards the pressure bump. This reduces the amount of CAIs passing through and thus lowers the CAI abundance in the region in between the two planets. A rapid transition seems to occur between $\alpha_\text{peak} = 1*10^{-4}$ and $2*10^{-4}$.}\label{fig:SaturnParameterSearch4}
\end{figure}

To answer this question, the model was extended with a second planet gap caused by the formation and growth of Saturn. We assume that Saturn has the same mass as Jupiter ($M = 30 M_{\oplus}$) when it starts opening a gap, and grows to its full size in the same amount of time, meaning it reaches a mass of $M = 95.2$ $M_{\oplus}$, or one Saturn mass, after 4.5 Myr. This leaves three important parameters for which we must choose some value: the formation time $t_\text{planet}$, the formation location $a_\text{planet}$, and the depth of the planet gap, represented by $\alpha_\text{peak}$ in Equation (42). Concerning the planet formation time, we assume that Saturn started forming around the same time as Jupiter ($t_\text{planet} = 0.6$ Myr). We initially place Saturn at its present-day location at $a_\text{planet} = 9.6$ AU, although it would make sense to place its birthplace closer to the Sun, since Jupiter also formed $\pm$ 40\% closer in in our model. The largest uncertainty lies in the value of $\alpha_\text{peak}$. As Saturn is less massive than Jupiter, it makes sense that its planet gap would not be as deep and its potential for trapping dust species in a pressure maximum less strong. This places an upper limit of 10$^{-2}$ on $\alpha$. A lower limit of 10$^{-5}$ is set by the viscosity in the vicinity of Saturn, as this is the lowest value of $\alpha$ in the surrounding disk, and any lower values would actually produce an overdensity instead of a gap. We rather arbitrarily assume a value of $\alpha_\text{peak} = 10^{-4}$ to start with. 

Figure \ref{fig:SaturnSigmaPop5} shows the evolution of the CAI surface density resulting from this model. Saturn's planet gap can be seen to the right of that due to Jupiter. While this gap is clearly less deep, a significant build-up of CAI surface density can still be seen behind it. However, this does not seem to prevent a similar build-up of CAIs in the region in between the two planets. This is reflected in Figure \ref{fig:SaturnCAIAbundance} showing the CAI abundance profile. Up until the formation time of both planets at 0.6 Myr, this result is identical to the one-planet case of Figure \ref{fig:CAIAbundance}. The inward drift of the second abundance peak where CAIs are piling up is also the same in this case, since this occurs independently from whether a planet is present or not. But in the meantime, the CAI abundance is rising faster at the location of Saturn's pressure bump than at Jupiter's. Saturn is stopping the inward drift of a significant fraction of the CAIs, but a large enough amount is still leaking through the gap to ensure that the final CAI abundance behind Jupiter is a significant fraction of what it is in the one-planet case. 

As with the main model, we can investigate how the results depend on the parameter choices. Figure \ref{fig:SaturnParameterSearch0} shows how the final abundances after 5 Myr depend on Saturn's formation time $t_\text{planet}$, while Jupiter's formation time is kept fixed at 0.6 Myr. Perhaps unsurprisingly, because it matches what happens in the one-planet case, the formation time has little effect on the end result. Sufficient CAIs remain in the disk, drifting inward, and it takes several million years for the second peak to drift all the way in, whether it be to 3 AU or 9.6 AU. Though this effect is not very strong, the CAI abundance in between the two planets does increase for later formation times for Saturn, as more CAIs can drift past its location in the meantime.  

Figure \ref{fig:SaturnParameterSearch1} shows how the results depend on Saturn's formation location $a_\text{planet}$, while keeping Jupiter fixed at 3 AU. Unlike in the one-planet scenario, this parameter now does have a rather large effect on the final abundance profile. Since the width of the planet gap depends on Saturn's Hill radius, which is proportional to its heliocentric distance, it shrinks as Saturn moves closer in. We have already seen that Saturn's pressure bump is less effective at keeping CAIs in place than Jupiter's is, and moving the formation location closer to the Sun exacerbates this issue. The closer in Saturn forms, the higher the CAI abundance in Jupiter's pressure bump becomes. The blue line for $a_\text{planet} = 5.6$ AU places Saturn $\pm$ 40\% closer in than where it is located today, similar to Jupiter. The result here greatly resembles the one-planet case with only a small dip at the location of the second planet gap. 

Finally, and quite as expected, Figure \ref{fig:SaturnParameterSearch4} shows that increasing the value of $\alpha_\text{peak}$, which causes a deeper gap and thus a stronger pressure gradient in the gas, which pushes CAIs back out towards the pressure maximum, will lead to higher CAI abundances in Saturn's own pressure bump, while letting through less CAIs in the direction of Jupiter. The difference in abundance at Jupiter's pressure bump seems particularly large between the cases with $\alpha_\text{peak} = 1\times10^{-4}$ and $2\times10^{-4}$, suggesting that perhaps there is a transition point around these values above which Saturn's CAI trapping capability changes from ineffective to effective.

Without more precise knowledge about the location where Saturn originally formed or what value of $\alpha_\text{peak}$ best represents how well it can push CAIs back towards its pressure bump, it is difficult to say what exactly the effect of a second planet on the CAI abundances in the Solar System would be. However, as we have demonstrated, the presence of multiple planet gaps in the solar protoplanetary disk would at least not necessarily be an impediment to obtaining the results of our main model as shown in Figure \ref{fig:CAIAbundance}. The inclusion of the infall phase from a collapsing molecular cloud into the model by Desch et al (2018) therefore does not transport CAIs out too far for the Jupiter solution to still work as an explanation for the relatively high CAI content of carbonaceous chondrites.

\subsection{A gravitationally stable model}\label{A Gravitationally Stable Model}

\begin{figure*}[t!]
\minipage[t]{\columnwidth}
  \includegraphics[width=\columnwidth]{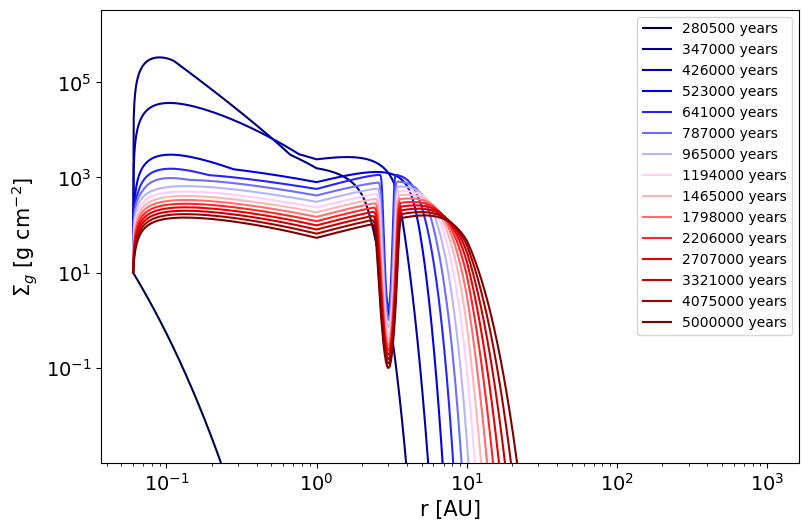}
  \caption{Viscous evolution of the gas surface density $\Sigma_g$ for a model with a reduced cloud rotation rate and no gravitational instability.}\label{fig:StableSigma}
\endminipage\hfill
\minipage[t]{\columnwidth}
  \includegraphics[width=\columnwidth]{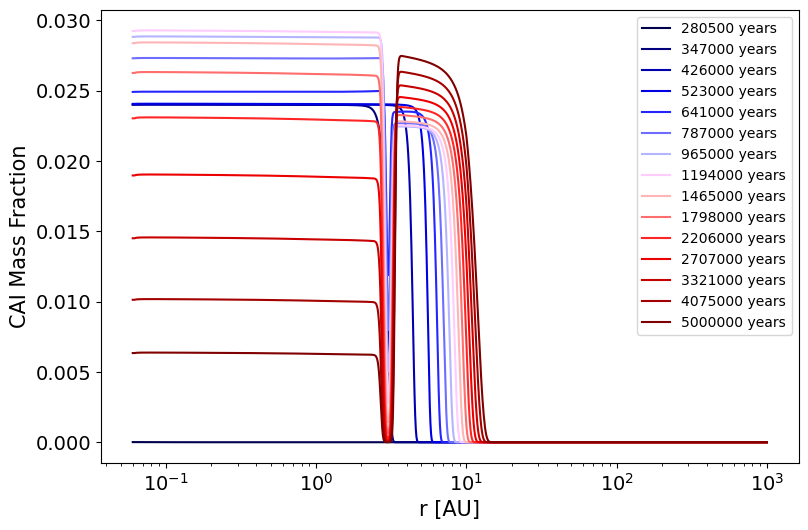}
  \caption{Time evolution of the CAI abundance throughout the disk for a model with a reduced cloud rotation rate and no gravitational instability.}\label{fig:StableCAIAbundance}
\endminipage\hfill
\end{figure*}

\begin{figure*}[h!]
\minipage[t]{\columnwidth}
  \includegraphics[width=\columnwidth]{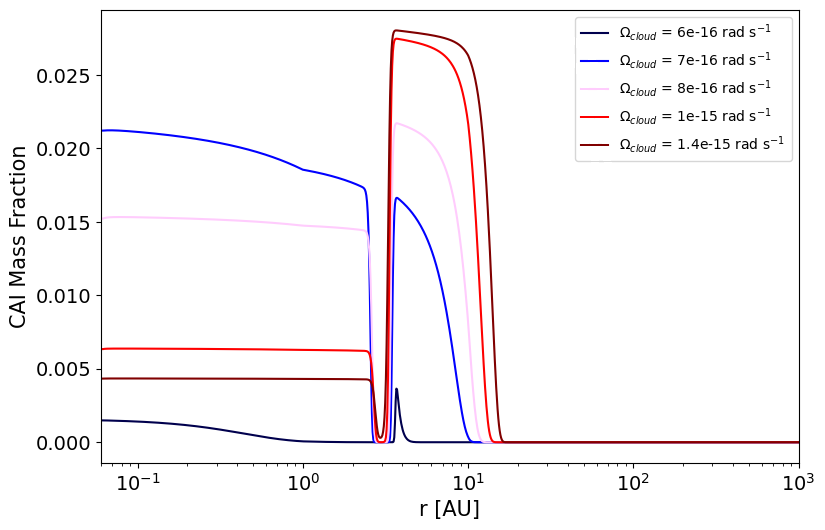}
  \caption{Abundances after 5 Myr for different values of the cloud rotation rate $\Omega_c$. There is only a narrow range of values for which the abundance profile agrees with meteoritic observations without triggering the gravitational instability.}\label{fig:StableParameterSearchomcloud}
\endminipage\hfill
\minipage[t]{\columnwidth}
  \includegraphics[width=\columnwidth]{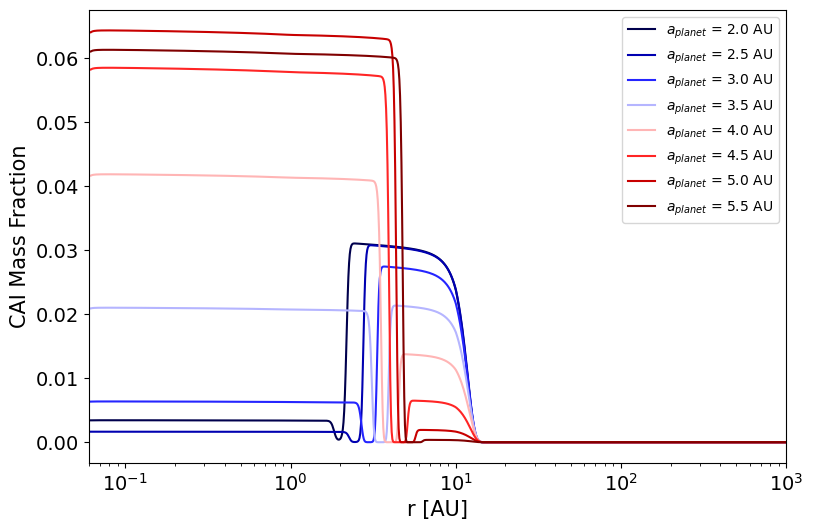}
  \caption{Abundances after 5 Myr for different values of the planet formation location $a_\text{planet}$. For formation locations at greater radial distances than 3 AU, the CAI abundance behind the planet gap rapidly decreases, while it increases in the inner disk.}\label{fig:StableParameterSearchaplanet}
\endminipage\hfill
\end{figure*}

Up until now, we have, through the choice of values for the molecular cloud rotation rate and temperature, assumed that so much mass is deposited onto the protoplanetary disk that it becomes gravitationally unstable. It is however possible that the Sun formed in a region of high-mass star formation, where temperatures tend to be higher and strong magnetic fields can slow down the cloud rotation rate through magnetic braking. This leads to clouds with lower angular momentum, and therefore also smaller centrifugal radii during disk formation. In such a situation it is possible that the disk never becomes gravitationally unstable. To see how this affects disk (and CAI) evolution, we ran a model with lower $\Omega_c$, keeping our other disk parameters the same as before. We found that $Q_\text{Toomre}$ exceeds 2 in at least some part of the disk for rotation rates $\Omega_c$ of at least $2\times10^{-15}$ rad s$^{-1}$, and thus picked a value of $\Omega_c = 1\times10^{-15}$ rad s$^{-1}$ for this new simulation. Since no artificially increased $\alpha$-parameter is required in this model, we use the parameterization from Equations (36-38) from the start.

Figure \ref{fig:StableSigma} shows the resulting time evolution of the gas in this disk model. Comparing this to Figure \ref{fig:SigmaGas}, there are several notable differences. First of all, the disk now only starts forming after 0.28 Myr, as the centrifugal radius grows more slowly and requires more time to even reach the inner disk edge at 0.06 AU. Because the infall phase still ends after 0.4 Myr, disk formation now occurs in a shorter time span. Secondly, the radial extent of the disk is much smaller. At the end of infall, the centrifugal radius is no larger than 0.18 AU, and without the gravitational instability pushing material outward, a sharp drop in the surface density can be observed near $r = 10$ AU. Although this disk ends up with less total mass than in the gravitationally unstable case, this mass is much more concentrated in the inner disk. Therefore the third major difference is that surface densities in the inner disk are now 1-2 orders of magnitude higher. Since this increases the viscous heating rate, the radial extent of the CAI Factory has also increased, though only slightly, to 1 AU. 

Figure \ref{fig:StableCAIAbundance} shows the evolution of the CAI abundance profile in this model. The denser gas slows down inward radial drift of CAIs, and they remain spread out fairly evenly between the pressure bump and $r = 10$ AU, beyond which virtually no CAIs are present. The peak abundance value of nearly 3\% is naturally lower in this result than in the gravitationally unstable case, and in line with observed meteoritic abundances of a couple percent. In order to achieve a result where the inner disk drains of CAIs however, we had to increase the planet gap depth by increasing the value of $\alpha_\text{peak}$ in Equation (42) from 0.01 to 0.1. Without this adjustment, there is sufficient leakage of CAIs through the planet gap that the abundance on both sides remains almost the same. 

As Figure \ref{fig:StableParameterSearchomcloud} shows, there is only a narrow range of values for $\Omega_c$ for which the final abundance profile resembles that in Figure \ref{fig:StableCAIAbundance}. Increasing the rotation rate by a factor two leads to a situation in which the gravitational instability kicks in again, while the trapping effect rapidly becomes ineffective for lower values. When $\Omega_c = 6\times10^{-16}$ rad s$^{-1}$, the centrifugal radius barely exceeds the disk inner edge anymore, and very few CAIs are left in the disk to be trapped. We did not explore the effects of moving the disk inner edge closer to the Sun.

The final result we'll show for our gravitationally stable model is a parameter search over the planet formation location $a_\text{planet}$. The smaller radial extent of this disk increases the significance of this parameter considerably. As Figure \ref{fig:StableParameterSearchaplanet} shows, moving Jupiter's formation location out any further than 3 AU will rapidly decrease the CAI abundance in the pressure bump, and increase the abundance in the inner disk, as not enough CAIs get trapped behind its orbit. 

\section{Discussion}\label{Discussion}

The main science question we set out to answer was whether the DKA18 solution to the CAI storage problem still works when the effects of the infall phase from a parent molecular cloud core are included into the model. In a nutshell, we can answer this question positively. But while Jupiter's planet gap does serve as an effective barrier preventing CAIs from drifting into the inner solar system and eventually vanishing into the Sun, there are also some important differences between the models. 

\subsection{Main model}\label{Discussion Main Model}

The main focus of our work has been on a model in which the solar protoplanetary disk becomes gravitationally unstable during the infall phase, as we found that this situation arises for a wide range of input parameters. The first new result that we found in this model was that CAIs are only created during the infall phase when the disk is still building up. During this phase, the viscosity in the disk is so strong that viscous heating alone will cause an extended region in the inner disk to reach a temperature of 1400 K. This has a very important consequence: the rapid viscous expansion of the gas during the infall phase drags the various dust species along with it, spreading them out much further than would be achieved by turbulent diffusion or meridional flow in an already established disk would. In the DKA18 model, there are virtually no CAIs present at a heliocentric distance of 10 AU, while we find them to exist even a hundred times further out than that. 

The fact that CAIs are transported out so far into the outer disk leads to another important result. The required time for CAIs to drift all the way back in towards the Sun is now several million years, instead of the several tens of thousands of years it would take them to vanish from the disk when they are only found in the innermost part. The presence of CAIs in the disk at the time when meteorite parent bodies formed is thus a natural consequence of the viscous spreading of the infall phase. This solves the second part of the CAI storage problem without needing to invoke a planet. 

We briefly considered the possibility that the first part of the CAI storage problem might also be solved without Jupiter trapping any dust particles, due to the pile-up of CAIs creating a distinct peak in the outer disk that slowly drifts in. This second peak forms regardless of whether a planet exists in the model or not. However, without a planet in the model to keep CAIs in place, the inner disk will continuously be seeded with new dust grains drifting in, and the CAI abundance remains very high in the inner disk. So while Jupiter is not needed to explain the presence of CAIs beyond its orbit, it is still required in order to let the inner disk drain successfully.

Another important difference between our model and the DKA18 model is the width of the abundance peak at the end of the simulation. In the DKA18 model, this peak extends to roughly 4 AU. In our model, it extends all the way out to 100 AU, although it must be noted that this is not entirely because of Jupiter's pressure bump. The inward radial drift of CAIs is a still ongoing process after 5 Myr, that would presumably continue if the disk itself would not dissipate yet. 

This model therefore predicts that CAIs should also be present in objects originating from far beyond the orbit of Jupiter, such as for example in Kuiper Belt objects. 

A final important difference we found is that in our model, the CAI abundance is significantly higher in the particle trap than in both the DKA18 model and real meteorites. While this seems problematic, the overall CAI abundance could be lowered by different parameter choices, most notably for the molecular cloud parameters, or by perhaps considering more realistic size distributions of CAIs.

There are also some notable similarities between our model and that of Morbidelli et al (2022). Their model also begins with infall from a molecular cloud core. The early, high-viscosity disk undergoes rapid viscous expansion as the radial velocity of gas is positive beyond the centrifugal radius, efficiently transporting dust outward. As in our case, the disk then evolves into a low-viscosity accretion disk as infall ends. Their model then manages to predict the contemporaneous formation through the streaming instability of two isotopically distinct planetesimal groups, one at the snowline around 5 AU and another at the silicate sublimation line around 1 AU. These groups correspond to the parent bodies of CC and NC meteorites, respectively. The difference in composition is caused by a change over time in the composition of the molecular cloud core (which we did not include in our model). Planetesimals at the snowline incorporate mostly material that accreted onto the disk early on, transported outward during the expansion phase, while later infalling material changed the overall composition of the dust in the inner disk. They note, but did not model, that a barrier against dust drift, likely in the form of Jupiter's formation, is still required to prevent the disk from homogenizing before later planetesimal formation is complete. CAIs in their model are assumed to have condensated from early infalling material. 

\subsection{Gravitationally stable model}\label{Discussion Gravitationally Stable Model}

An argument against CAIs spreading out into the (far) outer disk early on is the observed lack of CAIs in CI chondrites. Only one CAI has been found in a CI chondrite (Frank et al. 2011). CI chondrite parent bodies are thought to have formed after 3-4 Myr at heliocentric distances $r \geq 15$ AU (Desch et al. 2018), implying that CAIs had not reached these distances in significant quantities yet at this time. This observational constraint can be satisfied by not triggering the gravitational instability, which requires a molecular cloud with less angular momentum than in our main model. This can be achieved by considering models with lower cloud rotation rates and/or higher temperatures. These conditions can be achieved in high-mass star formation regions, where cloud cores with strong magnetic fields can slow down their rotation through magnetic braking (Wurster \& Li 2018). There is evidence that the Sun may have formed in such an environment (Hester \& Desch 2005). By using a smaller cloud rotation rate, we were able to produce a gravitationally stable model in which CAIs do remain trapped behind the planet gap, but don't extend out as far as 15 AU.\footnote{We did not explore different combinations of the cloud rotation rate and temperature that might yield similar results.} However, while this matches the observation that (almost) no CAIs are found in CI chondrites, it does not explain why they are found in comets that formed in the same region but at later times. Our current models only predict that CAIs were transported to this region early (with gravitational instability) or not at all (without it). 

\subsection{Suggested improvements}\label{Suggested Improvements}

There are several ways in which the model could be further improved. It might also be good to check the results of the model with a full hydrodynamic code.

The midplane temperature calculation we made use of could be improved upon. The model used for the stellar luminosity (Baraffe et al. 2002) is not really reliable for the earliest phases of the simulation, where an accurate temperature calculation is arguably the most important. We also assumed a constant opacity throughout the disk, instead of calculating it from the available dust components. A more precise temperature model might influence when and where exactly the CAI Factory is active.

We have seen that, at least for the $\alpha$-parameterization we employed, the disk in our main model becomes gravitationally unstable during the infall phase. Because this can lead to the emergence of overdense clumps and spiral arms which can transport angular momentum outward, these effects can (at least crudely) be mimicked by artificially increasing the $\alpha$-parameter for the viscosity. A full treatment of the gravitational instability would require a code that can handle non-axially symmetric disk models however.

The way in which the planet gap is introduced into the model is also quite crude. While the width of the gap grows with the planet's mass, its depth does not. This is an important parameter however, because it determines how well the particle trapping effect works. A more sophisticated model could be used for the opening of the gap. Also not taken into account is the possibility that the gap itself moves over time due to migration of Jupiter.

The disk mass in our main model evolves very little after the end of the infall phase, when most of the mass resides at large heliocentric distances, but the viscosity is too weak to move much of it back in to accrete onto the Sun. This is probably not very realistic. The model of Yang and Ciesla (2012) suffered from the same issue. Perhaps an $\alpha$-profile could be set up in such a way that a stronger accretion rate emerges, or different mechanisms of angular momentum transport could be included, such as magnetized winds. The dissipation phase of the disk due to photoevaporation could also be included into the model. This might at the same time also impact surface densities and radial drift velocities when meteorite parent body formation is still ongoing.

We have only modelled a single size of CAI grains simultaneously, even though CAIs exist in a size range from microns up to a centimeter. While we checked in section \ref{Parameter Search} how the final CAI abundances change for grains of different sizes (see Figure \ref{fig:ParameterSearch14}), each of these iterations still assumes that only that particular size of grain exists. A more realistic model would include a more complete sample of the CAI size distribution. This could be complicated to achieve however, because it is not clear whether CAIs of different sizes are created at the same time or with the same efficiency. It could equally well be that mostly large CAIs are created in the CAI Factory, with smaller CAIs being the result of later fragmentation in the disk. 

The effects of photoevaporation could be especially important in high-mass star formation environments with a high FUV flux, as this could have a significant impact on the disk evolution, for example by truncating the disk through rapid mass loss at the outer edge. It is also a possibility that mass loss due to photoevaporation might keep the disk gravitationally stable for higher values of the cloud rotation rate $\Omega_c$.

Finally, something that has also not been explored in this project is the possibility of a non-homogeneous distribution of matter in the molecular cloud, or a composition that changes over time. 

\section{Conclusion}\label{Conclusion}

The model we built shows that the solution that Desch et al (2018) proposed for the CAI storage problem, in which a pressure maximum created by Jupiter opening a gap in the disk traps CAIs in place, also works when taking into account that the solar protoplanetary disk formed out of a collapsing molecular cloud core. We find that CAIs are created during the infall phase and are then very efficiently transported outward by the combined effects of advection by the rapidly expanding gas, redistribution of matter due to possible gravitational instability and turbulent diffusion. 

Our main focus was on a disk model massive enough to become gravitationally unstable. In this case, subsequent inward radial drift creates a double peak structure in the CAI abundance. As well as piling up in Jupiter's pressure bump, CAIs in the far outer disk start piling up when they drift into a region of higher gas surface density, decreasing their Stokes number and slowing them down. The two abundance peaks keep growing over the course of the simulation until eventually merging together, forming a broad ($\pm$100 AU) region with elevated CAI abundances starting just behind Jupiter, where carbonaceous chondrites are then assumed to have formed. An interesting result from this extended period of inward radial drift is that the presence of CAIs in the disk after 4-5 Myr is a natural consequence of the infall phase, and does not actually require Jupiter as an explanation. In the meantime, the inner disk drains of CAIs as they drift into the Sun, leaving a much lower level of abundances in the region where ordinary and enstatite chondrites then form. 

For input parameters that do not lead to the gravitational instability, the final CAI abundance profile in the disk looks qualitatively similar. The abundance peak is less wide however, as the disk is naturally smaller, and no double peak structure is observed.

We find that the results of our model do not strongly depend on most parameter choices. This applies even to some parameters that the DKA18 model is more sensitive to, such as (in the gravitationally unstable case) the time and location where Jupiter forms. The molecular cloud properties seem to have the largest quantitative impact on the CAI abundance. More importantly, the general shape of the CAI abundance profile is not disrupted by most parameter variations, so the Jupiter solution works for any sufficiently large dust particle. 

The presence of multiple planet gaps due to the other gas planets in our Solar System is not necessarily an impediment for the CAIs to drift back in towards Jupiter, as we showed that there are at least reasonable parameter choices that cause additional planet gaps to let through significant amounts of CAIs.

Quantitatively, the CAI abundances our model predicts are quite uncertain, not only due to parameter uncertainty, but also due to simplifications in the model, such as the singular CAI grain size. Qualitatively however, we are confident that the described results are authentic. 

\begin{acknowledgements}

We thank the referee, Steve Desch, for an extensive and very valuable referee report which greatly helped us to improve the paper.
\newline\newline
This research was supported by the International Space Science Institute (ISSI) in Bern, through the ISSI International Team project "Zooming In On Rocky Planet Formation".

\end{acknowledgements}

\begin{appendix}

\section{Choice of time steps}

During the infall phase, when the value of the $\alpha$-parameter is artificially increased, mixing of gas and dust is particularly strong. Since the thermal conversion of dust species is not included in the implicit integration scheme, but solely occurs after each discrete time step $\Delta$\textit{t}, we must be careful to choose a time step small enough so that no significant amounts of dust can mix into the CAI Factory and out of it again in between two time steps. In such a scenario, the population 1 dust would not be processed into CAIs in our simulation, when it would have been in reality. 

\begin{figure}[h!]
  \centering
  \includegraphics[width=\columnwidth]{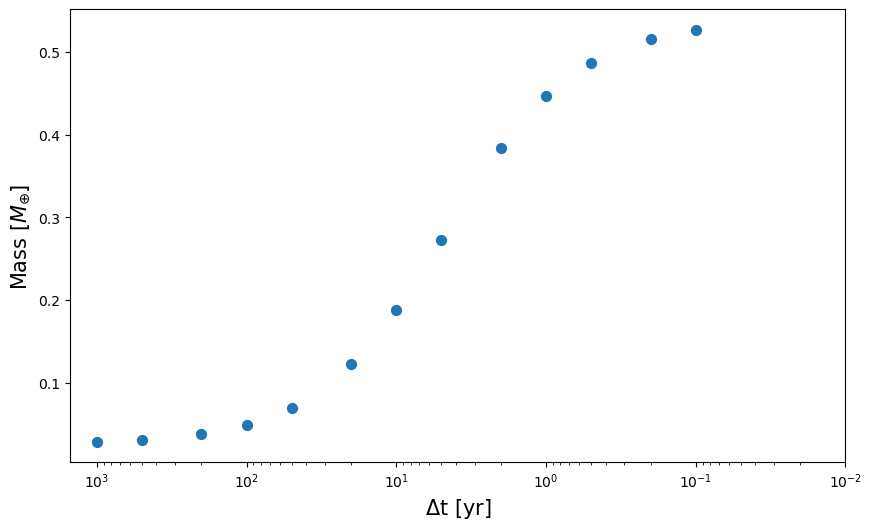}
\caption{Total CAI mass in our model after 0.1 Myr as a function of the time step $\Delta$\textit{t}. }\label{fig:ConvergenceTest}
\end{figure}

In order to determine a suitable time step, we ran our main simulation with a number of different time steps, ranging from 1000 down to 0.1 years. After 0.1 Myr, we stopped the simulation and checked the total CAI mass present in the model at that point in time. The results are plotted as Figure \ref{fig:ConvergenceTest}. 

The total CAI mass in the disk clearly starts to increase when decreasing the time step from the initial 1000 years, as less and less population 1 dust manages to enter and exit the CAI Factory in between two time steps, thereby avoiding conversion. We decided to use a time step of $\Delta$\textit{t} $= 0.2$ years for our simulations, as the CAI mass has more or less converged by this point, while still allowing all our simulations to be completed in a reasonable amount of time. 

After the conclusion of the infall phase, the $\alpha$-parameter is reduced to that in Equations (41-43), and gas and dust mixing occurs much less rapidly. A constant time step of 1000 years is then used for the remainder of the simulation. We find that the results would not significantly change had we used smaller time steps than this.

\end{appendix}

\end{document}